\documentclass[twocolumn,showpacs,preprintnumbers,amsmath,amssymb]{revtex4}

\usepackage{graphicx}
\usepackage{dcolumn}
\usepackage{bm}
\usepackage{dsfont}
\usepackage{mathrsfs}
\usepackage{amsmath}				
 
\usepackage[bookmarks,
                 bookmarksopen = true,
                 bookmarksnumbered = true,
                 linktocpage,
                 colorlinks = true,
                 linkcolor = blue,
                 urlcolor  = blue,
                 citecolor = blue,
                 anchorcolor = green,
                 hyperindex = true,
                 hyperfigures]
		{hyperref}

\usepackage{color}

\newcommand{\be}{\begin{equation}}
\newcommand{\ee}{\end{equation}}
\newcommand{\bey}{\begin{eqnarray}}
\newcommand{\eey}{\end{eqnarray}}

\begin{document} 
\title{Fractality and other properties of center domains at finite temperature\\
Part 1: SU(3) lattice gauge theory} 
\author{ 
Gergely Endr\H{o}di$^{\,a}$, Christof Gattringer$^{\,b}$, 
Hans-Peter Schadler$^{\,b}$}
\affiliation{
\vspace{3mm}
$^a\,$Institut f\"ur Theoretische Physik, 
Universit\"at Regensburg, 94040 Regensburg, Germany \\
$^b\,$Institut f\"ur Physik,
Karl-Franzens-Universit\"at, 8010 Graz, Austria
}

\date{January 30, 2014}

\begin{abstract}
\vspace{3mm}
Using finite temperature SU(3) lattice gauge theory in the fixed scale approach
we analyze center properties of the local Polyakov loop $L(x)$.  We construct spatial clusters of 
points $x$ where the phase of $L(x)$ is near the same center element and study their 
properties as a function of temperature. We find that below the deconfinement transition
the clusters form objects with a fractal dimension $D < 3$.
As the temperature is increased, the largest cluster starts to percolate and its 
dimensionality approaches $D=3$.
The fractal structure of the clusters in the transition region may have implications regarding both the 
small shear viscosity and the large opacity of the Quark Gluon Plasma 
observed in heavy-ion collision experiments.
\end{abstract}


\maketitle
 
\noindent 
\section{Introductory remarks}
Various experiments at CERN, Brookhaven, GSI and NICA aim at studying Quantum Chromodynamics
(QCD) under extreme conditions. In particular analyzing and understanding properties of the
Quark Gluon Plasma (QGP) -- the high temperature phase of QCD where quarks are deconfined -- is
high up on the agenda. Several properties of the QGP could already be established
experimentally, such as its (almost) ideal fluid behavior or the large opacity, and the theory
side is challenged to contribute to the understanding of the underlying mechanisms. 

Some of the efforts for describing properties of the QGP~\cite{centermodels,Gupta:2010pp} have
revisited an old suggestion~\cite{centerdomains}, namely that center domains play a role in QCD
phenomenology. QCD without quarks, i.e., pure gluodynamics, has an interesting symmetry: A
transformation of the temporal component $A_4$ of the gluon field with an element of the
center $\mathds{Z}_3$ of the gauge group SU(3) leaves both the 
gluon action and the integration measure of the path integral invariant. However, at high
temperatures of roughly 300 MeV this center symmetry is broken spontaneously and a first
order transition leads to the deconfined phase. A suitable order parameter for this transition is the
Polyakov loop $L(x)$ defined as
\begin{equation}
L(x) = \mbox{Tr} \, {\cal P}  \exp \bigg( \int_0^{1/T} \! A_4(x,t) \, dt \bigg) \;,
\label{ploopdef}
\end{equation}
where $x$ is the spatial coordinate, ${\cal P}$ a path ordering operator, and the Euclidean 
time $t$ is integrated up to the inverse temperature. The Polyakov loop $L(x)$
corresponds to a static color source at $x$ and under a center rotation
transforms
non-trivially, $L(x) \longrightarrow z L(x)$ with  $z \in$ 
$\mathds{Z}_3 = \{e^{-i 2\pi/3}, 1, e^{i2\pi/3} \}$. Below $T_c \sim 300$ MeV the center symmetry
is intact and $\langle L(x) \rangle = 0$. The deconfined phase above $T_c$ is
characterized by a non-vanishing expectation
value $\langle L(x) \rangle \ne 0$. In this breaking of the center symmetry the phase of 
$\langle L(x) \rangle$ spontaneously selects one of the three elements $z \in$ $\mathds{Z}_3$.

Up to the fact that the ordered phase appears at high instead of low temperatures, the spontaneous
breaking of the center symmetry is equivalent to the temperature driven transition
of a 3-D ferromagnet with three possible spin orientations. Such a ferromagnetic
transition is often accompanied by the formation of domains where spins inside a domain
have the same orientation. In the interior of the domains the spins minimize their ferromagnetic
interaction and only the domain walls between domains of different spin orientation cost
energy. It is an interesting idea to explore whether such center domains could play a role
also in the deconfinement transition of gluodynamics
\cite{centerdomains}.

For full QCD, where the quark fields interact with gluons, the situation is slightly
different: The fermion determinant of the quarks explicitly breaks center symmetry and
the first order transition of pure gluodynamics is turned into a continuous crossover
near $T \sim 160$~MeV~\cite{crossover}. 
The fermion determinant favors the trivial center element $z = 1$, see, e.g., Ref~\cite{Kovacs:2008sc}.  
In the spin system language the fermion
determinant thus acts like an external magnetic field. However, the explicit breaking from
the quarks seems to be rather weak and center domains could still play a 
 role.

In this paper we aim at exploring structural properties of the center domains using lattice QCD
techniques. This approach has the advantage that the gauge field configurations which dominate
the path integral can be generated and it is straightforward to evaluate the Polyakov loop
$L(x)$ at each point $x$ in space. Thus possible center domains can  directly be inspected and
one may test whether characteristic properties are indeed such as assumed for explaining
features of the QGP. 

Here we consider the case of pure SU(3) lattice gauge theory where the Polyakov loop
is a true order parameter. In a second paper we will then look at the case of full QCD with 2+1
flavors of quarks. For the quenched case several studies can be found in the 
literature~\cite{quenched1,quenched1b,quenched2,quenched2b}, 
and preliminary results for full QCD are documented in
Ref.~\cite{full_prelim}. In all of them the temperature was driven by the inverse gauge coupling.
This means that temperature and cutoff effects are mixed which is problematic for analyzing 
some of the aspects of center domains, in particular the determination of the fractal structure
which we  study in this paper.  We here work in the fixed scale approach, i.e., the temperature
is driven by changing the temporal extent of the lattice. For the analysis of the scaling behavior we
compare results for three different lattice spacings. In addition we use different lattice
sizes to study finite volume effects.

\section{Properties of local Polyakov loops} 
We use configurations of pure SU(3) lattice gauge theory generated with the
Wilson action.  We work on lattices of sizes $N_s^3 \times N_t$ with $N_s = 30,
40$ and $48$ and with $N_t$ ranging from $N_t = 2$ up to $N_t = 20$. The
temperature is then given by $T = 1/N_t a$, where $a$ is the lattice spacing and
we have  set Boltzmann's constant to $k_B = 1$. Results are compared for three
values of the inverse gauge coupling, $\beta = 5.9, 6.2$ and $6.45$, which
correspond to lattice spacings of $a = 0.1117, 0.0677$ and $0.0481$ fm as determined
using Ref.~\cite{scale} with a Sommer parameter of $r_0 = 0.5$ fm.   For each set of
parameters we typically have 1000 configurations (except for our largest lattice)
and the error bars we show are statistical errors determined from a jackknife
analysis. Table \ref{parametertable}  summarizes the parameters of our runs. In
the last column we provide the value of the critical temporal extent $N_t^c$, such that
we can evaluate the temperature $T$ in units of the deconfinement temperature 
$T_c$ as $T/T_c = N_t^{c}/N_t$.

\begin{table}[b]
\caption{
Parameters of our ensembles. We list the inverse coupling 
$\beta$, the lattice spacing $a$ in fm, the size of the lattice, 
the number of configurations and the temporal lattice extent $N_t^c$ that 
corresponds to the deconfinement temperature.}
\label{parametertable}
\begin{center}
\begin{tabular}{cccrr}
\hline \hline
\quad $\beta\;\;$ \qquad & \quad $a$[fm] \qquad & \quad size \quad & 
  \# confs.  & \quad $N_t^c$ \quad \\
\hline 
5.90  & 0.1117 & \quad $30^3\! \times \! N_t$ & 1000 \quad & 5.978   \\
5.90  & 0.1117 & \quad $40^3\! \times \! N_t$ & 1000 \quad & 5.978   \\
5.90  & 0.1117 & \quad $48^3\! \times \! N_t$ & 1000 \quad & 5.978   \\
6.20  & 0.0677 & \quad $30^3\! \times \! N_t$ & 1000 \quad & 9.864  \\
6.20  & 0.0677 & \quad $40^3\! \times \! N_t$ & 1000 \quad & 9.864  \\
6.20  & 0.0677 & \quad $48^3\! \times \! N_t$ & 500  \quad & 9.864  \\
6.45 & 0.0481 &  \quad $30^3\! \times \! N_t$ & 1000 \quad & 13.884  \\
6.45 & 0.0481 &  \quad $40^3\! \times \! N_t$ & 1000 \quad & 13.884  \\
6.45 & 0.0481 &  \quad $48^3\! \times \! N_t$ & 500 \quad & 13.884  \\
\hline \hline
\end{tabular}
\end{center}
\end{table}

\begin{figure}[b]
\centering
\hspace*{-5mm}
\includegraphics[width=80mm,clip]{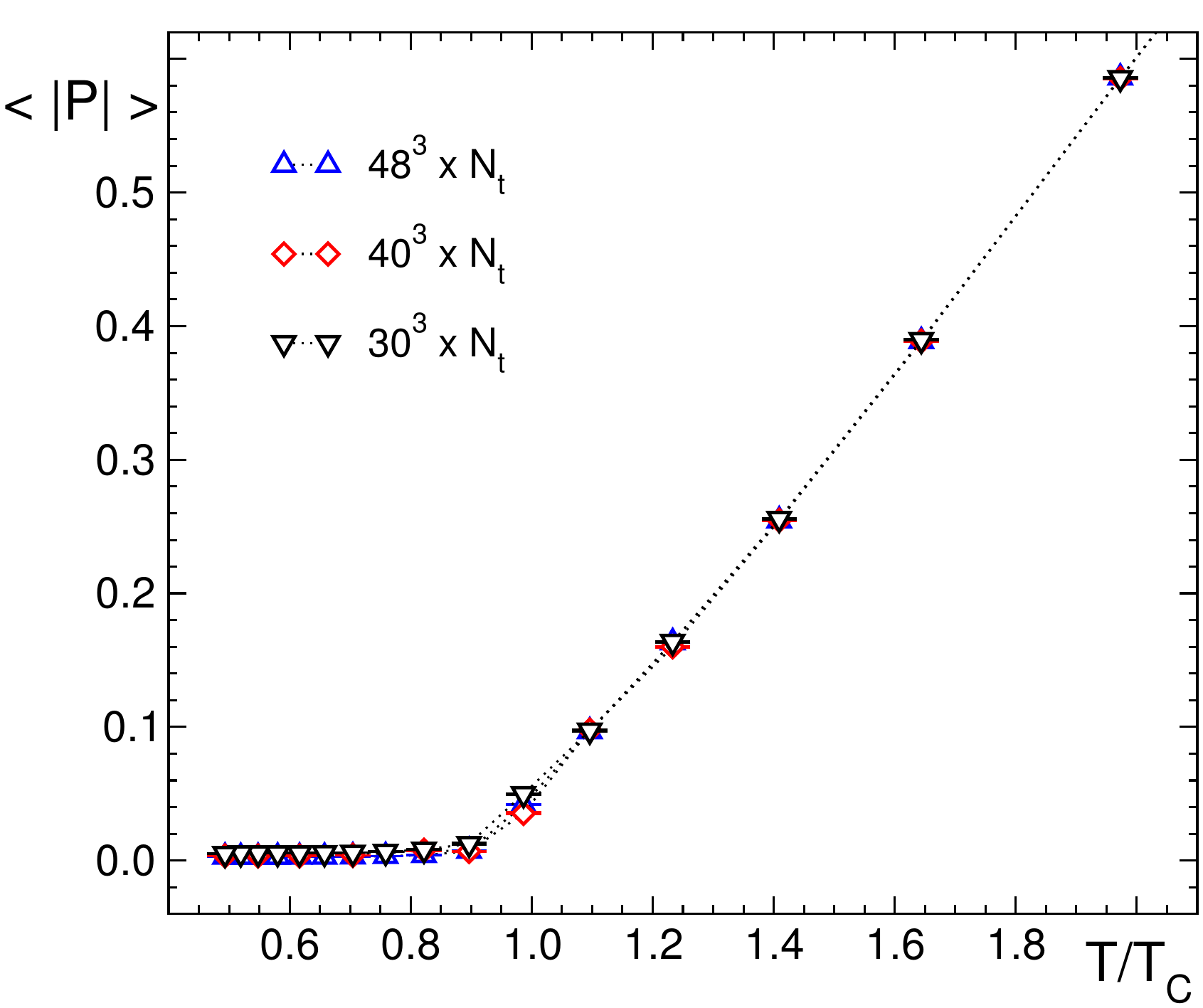}
\caption{The order parameter $\langle |P| \rangle$ as a function of the temperature. The results are
for our $\beta = 6.2$ ensembles.}
\label{Ploop_vs_T}
\end{figure}

For each gauge configuration we compute the local Polyakov loops $L(x)$ at all
spatial positions $x$. On the lattice $L(x)$ is given by the traced
product of all temporal links $U_4(x,t)$ at $x$,
\begin{equation}
L(x) \; = \; \mbox{Tr} \, \prod_{t=1}^{N_t} \, U_4(x,t) \; .
\label{Plooploc}
\end{equation}
As order parameter for center symmetry breaking one usually considers $\langle |P| \rangle$ where 
$P$ is the spatial average of $L(x)$, i.e.,
\begin{equation}
P \; = \; \frac{1}{N_s^3} \sum_x L(x) \; .
\label{Ploopaver}
\end{equation}

\begin{figure}[t]
\centering
\hspace*{-5mm}
\includegraphics[width=75mm,clip]{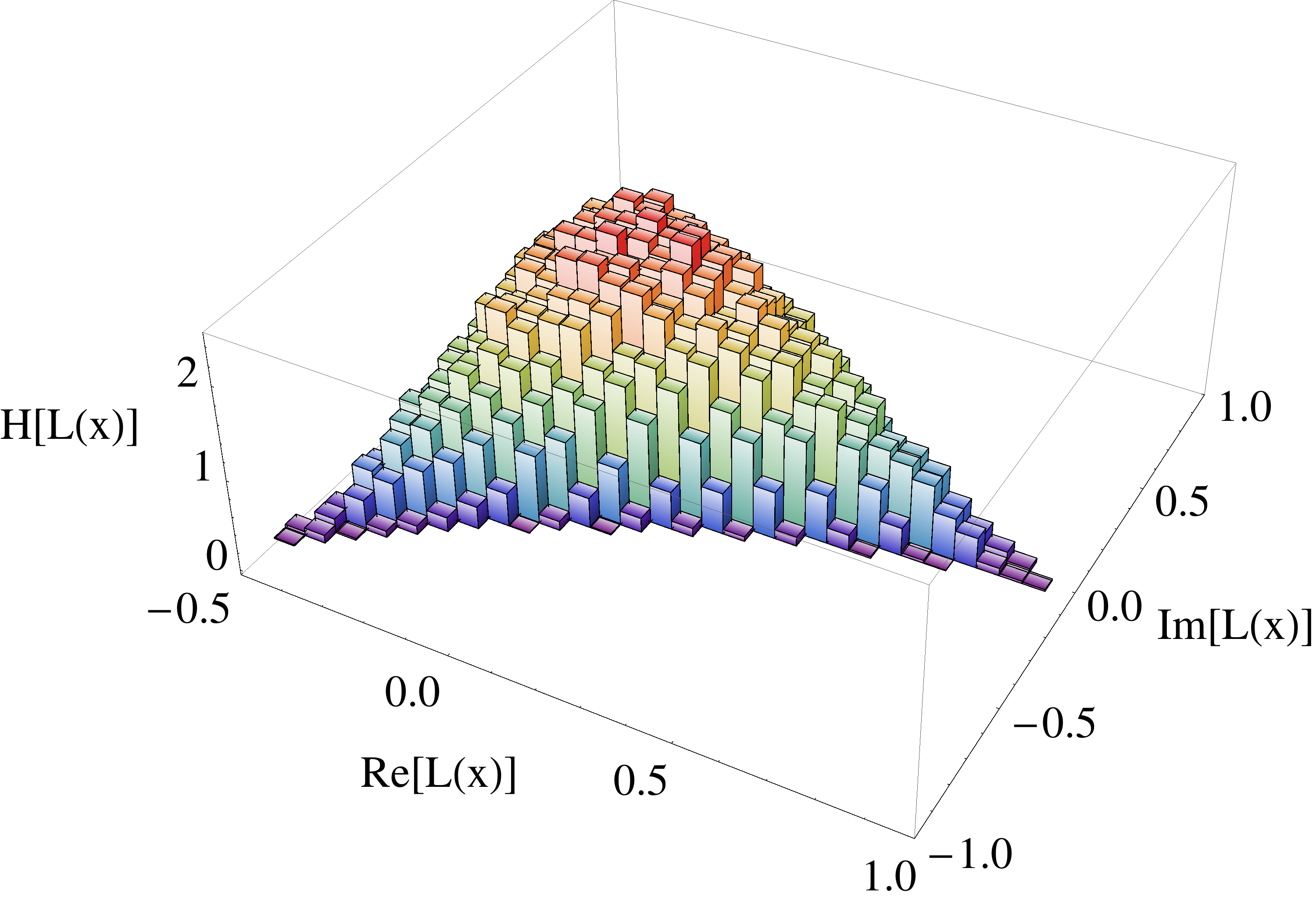}
\includegraphics[width=75mm,clip]{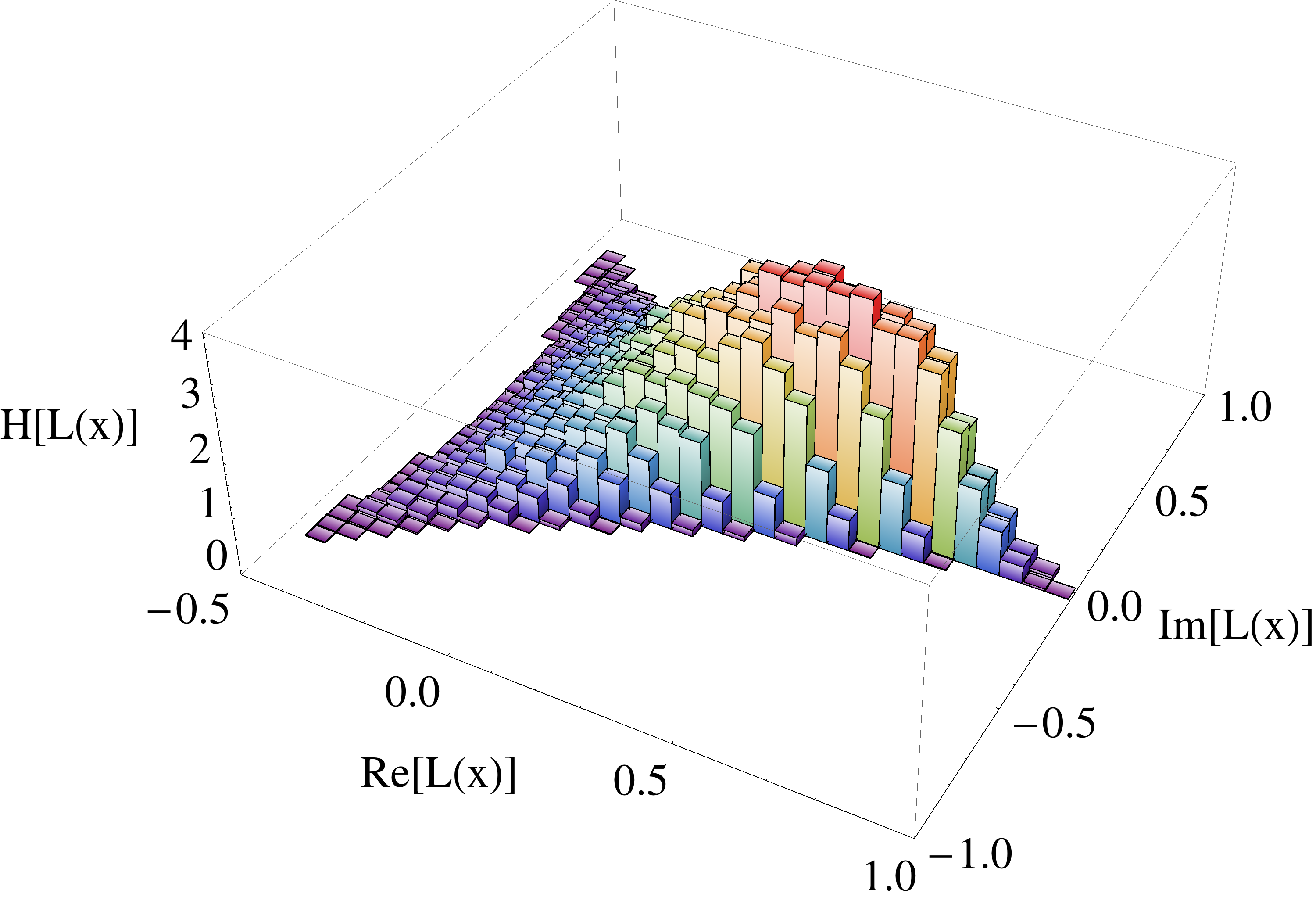}
\includegraphics[width=75mm,clip]{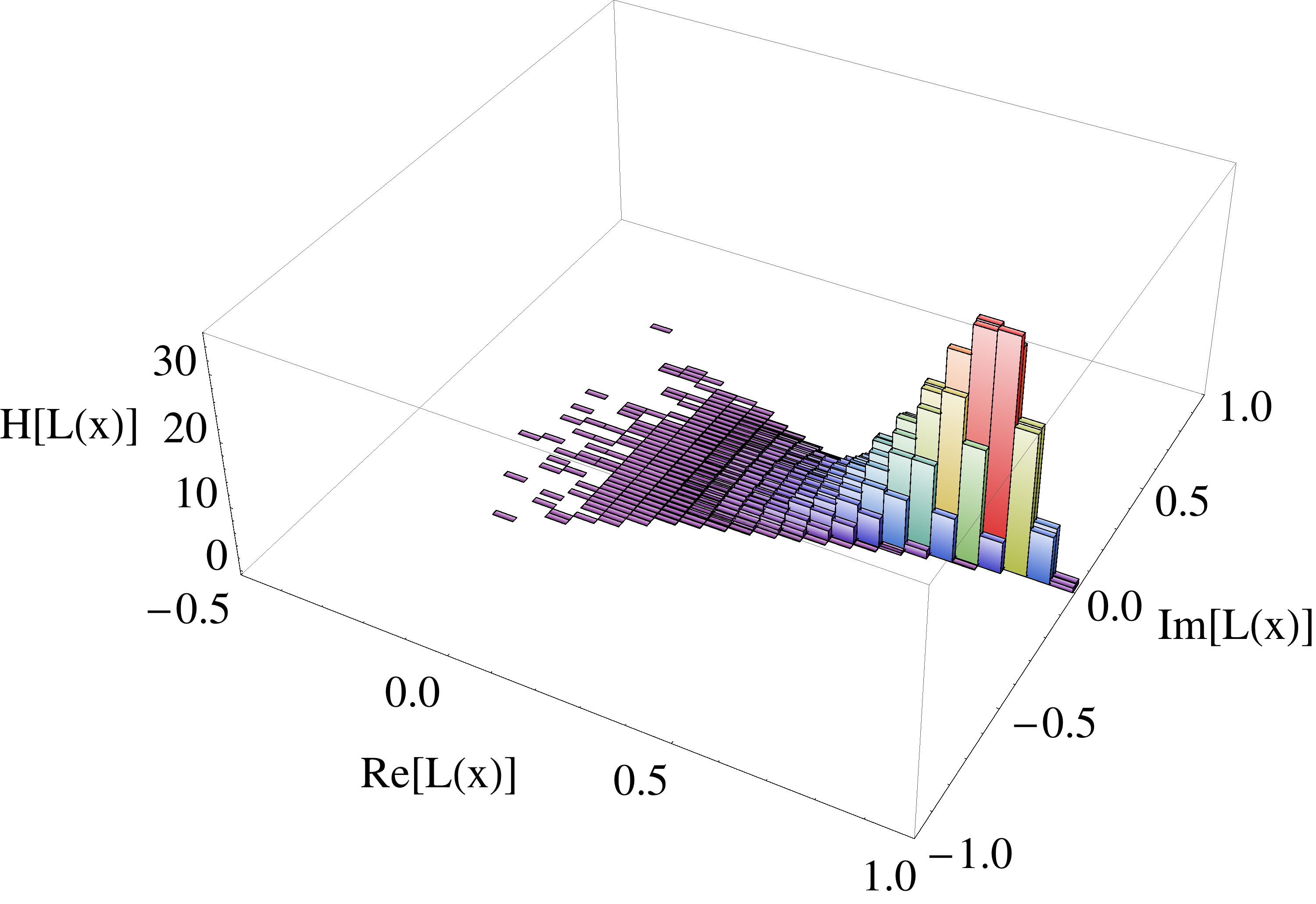}
\caption{Histograms for the distribution of local Polyakov loop values $L(x)$
in the complex plane. We show the results for 3 configurations from our 
$\beta = 6.2, 40^3 \times N_t$ ensembles, corresponding to temperatures 
$T = T_c$, $T = 2 T_c$ and $T = 5 T_c$ (top to bottom).}
\label{Ploop3D}
\end{figure}

For illustration purposes and later reference in Fig.~\ref{Ploop_vs_T} we show the results for $\langle |P| \rangle$ versus the temperature for our $\beta = 6.2, N_s=30, 40$ and $48$ 
ensembles. Below $T_c$ the Polyakov loop is zero, while after a small first order step
at $T_c$ it starts rising essentially linearly.

\begin{figure*}[t]
\centering
\hspace*{-5mm}
\includegraphics[width=150mm,clip]{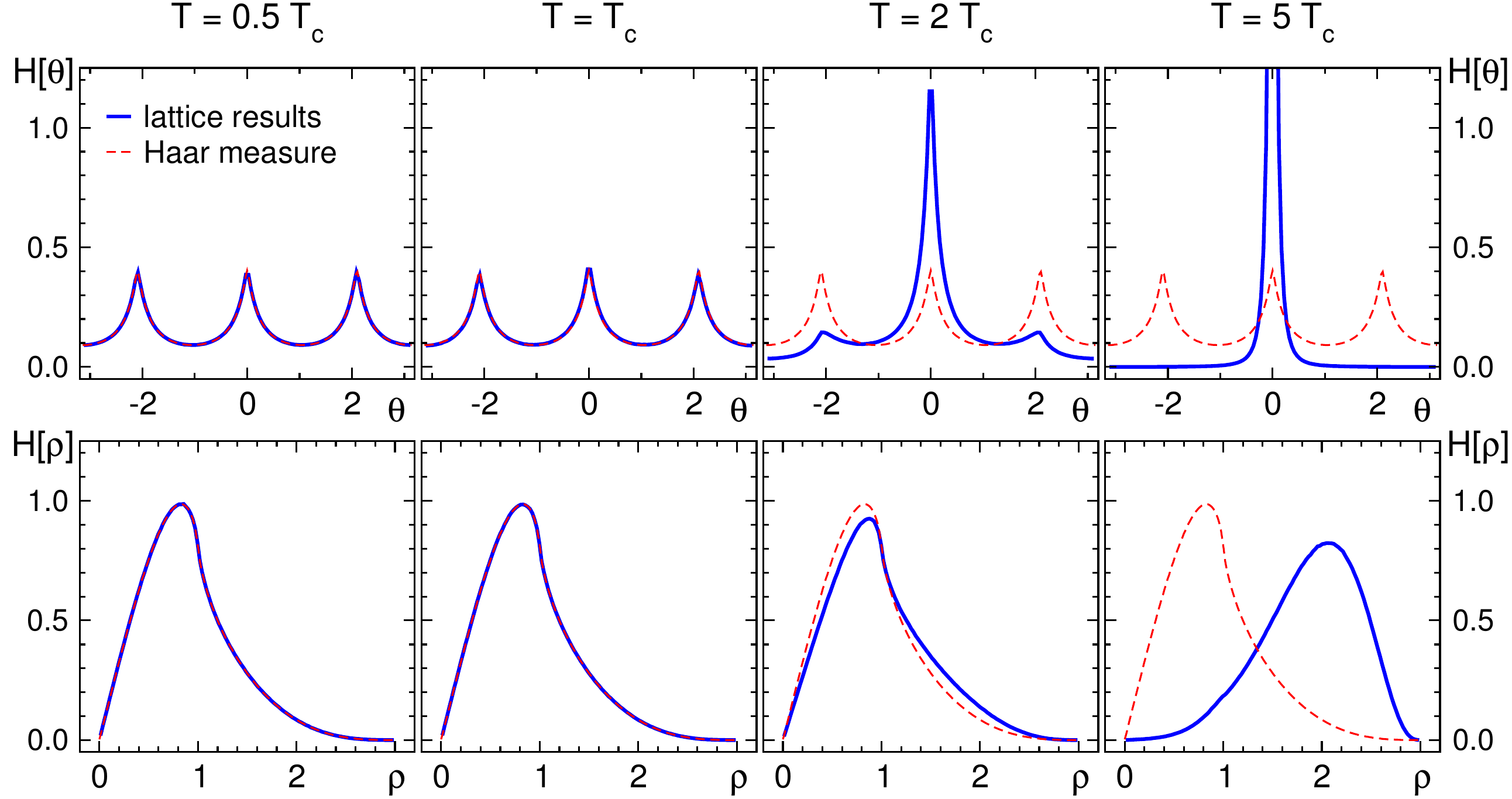}
\caption{Histograms of the phase $\theta(x)$ (top row of plots) and the modulus $\rho(x)$ (bottom) 
of the local Polyakov loops. We show results for four different temperatures and compare our Monte Carlo results (full curves) to the 
Haar measure distribution (dashed curves, which up to $T_c$ fall exactly on top of the 
Monte Carlo data). }
\label{histograms}
\end{figure*}

After having analyzed the behavior of the spatially averaged Polyakov loop $P$,
let us now come to  the local Polyakov loops $L(x)$. The local loops $L(x)$ are
complex numbers and in  Fig.~\ref{Ploop3D} we show histograms for their
distribution in the complex plane. Each plot is for a single configuration and we
show results at three different values of the temperature, $T = T_c$, $T = 2 T_c$ and $T = 5
T_c$.

For temperatures up to $T_c$ the values of $L(x)$ are distributed essentially
isotropically around the origin in the complex plane. In the spatially averaged
Polyakov loop $P$ of Eq.~(\ref{Ploopaver}), all $L(x)$ thus add up to zero
giving rise to $\langle |P| \rangle = 0$ below the deconfinement temperature
(compare Fig.~\ref{Ploop_vs_T}). Above $T_c$ the values of $L(x)$ are no longer
isotropically distributed, but start to show a preference towards one of the
center directions $1, e^{\pm i 2\pi/3}$. Thus in $P$ the $L(x)$ no longer
average to zero and $\langle |P| \rangle$ is finite above $T_c$.

For analyzing the distribution of the local Polyakov loops in more detail we
write them in polar form, 
\begin{equation} 
L(x) \;  =  \; \rho(x) \exp(i \theta(x)) \; , 
\end{equation} 
i.e., to every spatial  point we assign a modulus
$\rho(x)$ and a phase $\theta(x)$.

In Fig.~\ref{histograms} we show histograms for the distribution of the phase
$\theta(x)$ (top row of plots) and of the modulus $\rho(x)$ (bottom) for four
different temperatures (the histograms are for very fine bins, such that they
appear as continuous curves). We compare the results from the Monte Carlo simulation
(full curve) to the Haar measure distribution (dashed) given by \begin{equation}
P(\theta)  =  \!\int\! \!dU \delta( \theta - \mbox{arg Tr}\, U) \; , \;  P(\rho) 
=  \!\int\! \! dU \delta( \rho - | \mbox{Tr}\, U | )  , \end{equation} where $dU$
denotes the SU(3) Haar measure.

The plots show clearly that for $T \leq T_c$ the distribution of both the phase
$\theta(x)$ and the modulus $\rho(x)$ very closely follow the Haar measure
form. Above $T_c$ both distributions start to deviate from the Haar measure. 
At moderate temperatures above $T_c$ this deviation is considerably 
more pronounced for the
phases (compare also Ref.~\cite{quenched2}), and the distributions for the modulus start to show a
strong effect only for large temperatures ($T \gtrsim 3 T_c$). 

The decomposition of $L(x)$ into modulus and phase also sheds light on the
mechanism that leads to a vanishing spatially averaged Polyakov loop $P$ below 
$T_c$: Since the modulus is non-zero for most sites $x$, the mechanism must be predominantly
related to the phases. Below $T_c$ the distribution of phases is invariant 
under $\mathds{Z}_3$ center transformations, which shifts the distribution by $\pm 2 \pi/3$.
When averaging $L(x)$ over all spatial points $x$ the $\mathds{Z}_3$-symmetrically phases 
thus average to 0. Above $T_c$ the distribution of the phases is no longer 
$\mathds{Z}_3$-invariant and the system spontaneously (at infinite volume)
selects one of the three center sectors (the real sector in the examples shown
in our figures). The phases no longer average to zero when summing the $L(x)$
and $\langle |P| \rangle$ has a non-vanishing expectation value above $T_c$.

For the subsequent qualitative discussion we  will speak of a point $x$ being in a center
sector 0, $2\pi/3$ or $- 2\pi/3$ according to the closest  maximum in the distribution of
$\theta(x)$ shown in Fig.~\ref{histograms}. This preliminary definition of the center sectors
will be made more precise below.

For temperatures up to $T \sim 3 T_c$ the distributions of the local phase $\theta(x)$ in the top row plots of
Fig.~\ref{histograms} 
show peaks at all three center phases 0 and $\pm 2\pi/3$.
Up to $T_c$ 
these peaks have the same height and the distribution is center
symmetric, i.e., invariant under periodic shifts by $2\pi/3$. Furthermore, the distribution almost perfectly follows the 
Haar measure form (dashed curves). The fact that three peaks appear also above $T_c$ implies that also above the 
deconfinement transition 
a sizable fraction of the spatial points $x$ have a local Polyakov loop phase that is not in the dominant center 
sector the system has chosen in the spontaneous breaking of center symmetry. Only for rather high temperatures
$T > 3T_c$ all spatial points $x$ have settled on the same center phase $\theta(x)$.   

The interesting question is how the transition proceeds from a situation where all three center sectors are equally populated
(below $T_c$), through a situation where the majority of sites $x$ 
have already selected the dominant sector for the phase of $L(x)$, but a large fraction is still in a subdominant sector,
(between $T_c$ and 
$T \sim 3 T_c$) to the final state where all sites have chosen the same sector ($T > 3T_c$), and what phenomenological 
consequences this might have. In particular one may ask whether there are spatial domains or clusters where neighboring 
sites $x$ have their phases $\theta(x)$ pointing toward the same center element, and if such domains exist how their 
properties change with temperature.
 
 In order to study the formation and properties of center domains we now assign to the spatial points $x$ a center 
 sector number $n(x) \in \{-1,0,+1\}$ according to the following prescription:
\begin{equation}\label{eq:clusterdef}
n({x})  = \left\{ \begin{array}{rl}
-1 & \; \mbox{for} \;\; \theta({x}) \, \in \, 
[\,-\pi + \delta \; , \; -\pi/3 - \delta \, ] \; ,\\
0 & \; \mbox{for} \;\; \theta({x}) \, \in \, 
[\,-\pi/3 + \delta \, , \, \pi/3 - \delta \, ] \; ,\\
+1 & \; \mbox{for} \;\; \theta({x}) \, \in \, 
[\,\pi/3 + \delta \, , \, \pi - \delta \,] \; . 
\end{array} \right. 
\end{equation}
In our definition we introduce a real and non-negative cut $\delta$, which we write as
\begin{equation}
\delta \; = \; f \, \frac{\pi}{3} \; \; , \;\; \;\; f \in [0,1) \; .
\end{equation}
$f$ will be referred to as the cut parameter.
A value $f > 0$  has the effect that for sites $x$ where the phase $\theta(x)$ is far from each of the three 
center angles 0, $\pm 2\pi/3$ no center sector number is assigned. For $f = 0$ no sites are excluded, while for $f \rightarrow 1$ all
sites are removed from the analysis. 
 
In the subsequent construction of clusters these points are not taken into account and thus are removed from the analysis of the 
center domains. The introduction of the cut parameter $f$ allows us to consider only points $x$ where the phase of 
$L(x)$ clearly favors one of the center elements. Below we will analyze in detail how $f$ allows one to analyze clusters
at different length scales and discuss its physical role in more detail.

Spatial clusters of coherent center phase are now constructed according to the following prescription: 
Two neighboring points $x$ and $y$ are assigned to the same cluster if $n(x) = n(y)$. 
Standard cluster identification techniques can be applied to identify all clusters. 
Note that our definition of the clusters is insensitive to the renormalization of the 
Polyakov loop. Indeed, this renormalization amounts to a multiplication of $L(x)$ 
by a real, $x$-independent factor, i.e., it only changes the magnitude of $L(x)$. 
Since the clusters are 
defined in terms of the phases, they are not affected by the renormalization at all.

\begin{figure}[b]
\centering
\hspace*{-5mm}
\includegraphics[width=80mm,clip]{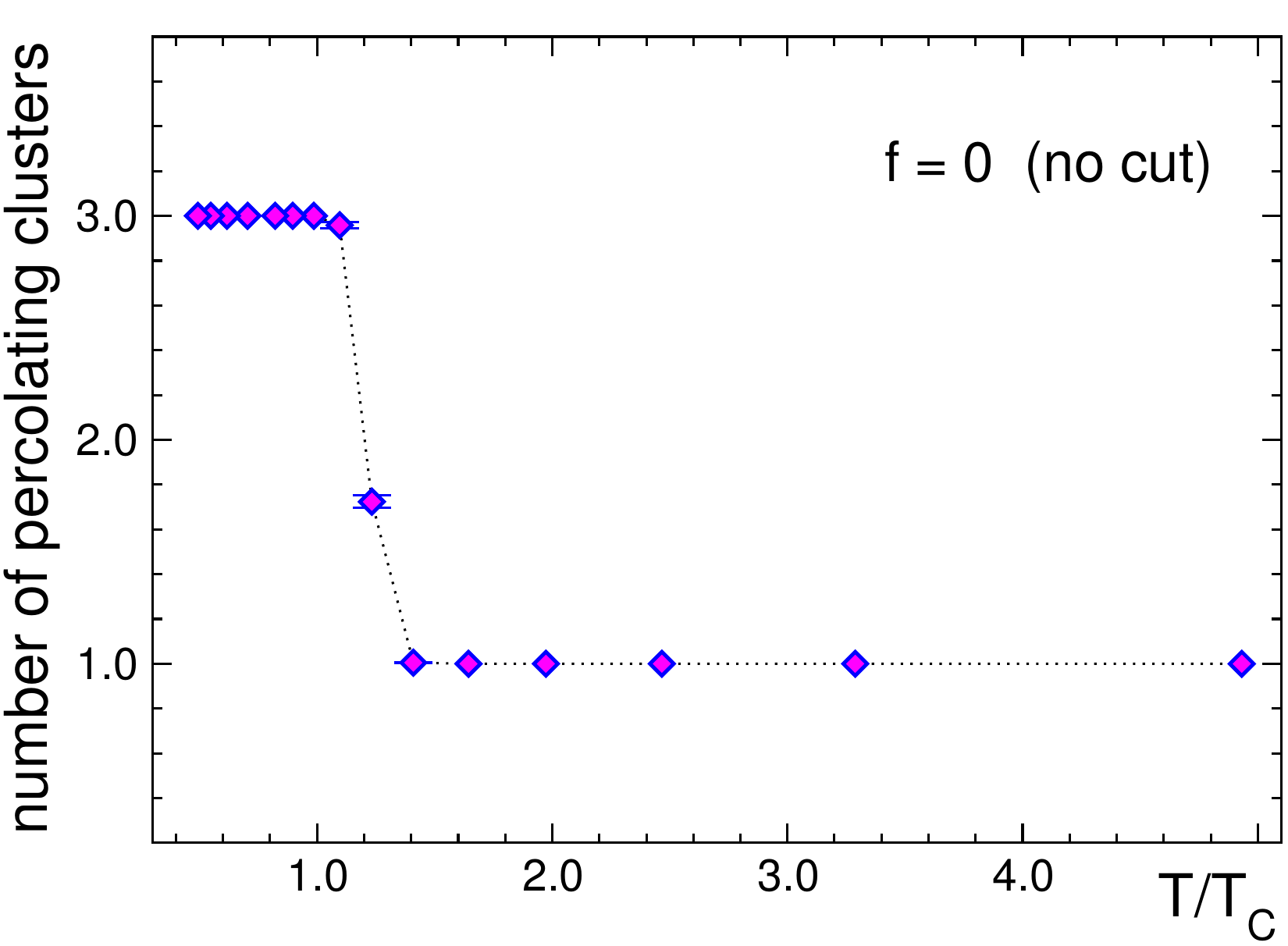}
\caption{The number of percolating clusters for cut parameter $f = 0$ (no cut) as a function of the temperature. The results are
for our $\beta = 6.2, 40^3 \times N_t$ ensembles.}
\label{f0clusters}
\end{figure}

Let us briefly recall an important fact from percolation theory:  For random site percolation
in 3 dimensions (i.e., all sites of a cubic lattice are  occupied with probability $p$ and
occupied neighbors are assigned to the same cluster) the critical probability where percolation
sets in is $p_c = 0.3116$~\cite{stauffer1994introduction,critperc}. Thus if the distribution of the  $n(x)$ was completely random, for the
case of $f = 0$ each site would be assigned $-1$, 0 or $+1$ with probability $1/3 > 0.3116$  and
we would always find percolating clusters for each of the center sectors, i.e., clusters that spread over all of the lattice 
(for the precise definition of percolation see below). 
Of course we do not expect a completely random distribution here, since the underlying dynamics correlates the center sector
numbers  $n(x), n(y)$ of neighboring sites $x, y$. 
 
In order to analyze this property, in Fig.~\ref{f0clusters} we show the number of percolating
clusters as a function of  temperature for the case of no cut, i.e., for $f = 0$. A percolating
cluster is defined as a cluster that has at least one site  in each of the  3 $N_s$ spatial
planes ($N_s$ $x_1$-$x_2$ planes, $N_s$ $x_2$-$x_3$ planes and  $N_s$ $x_1$-$x_3$ planes). The
results are for our $\beta = 6.2$ ensembles on $40^3 \times N_t$ lattices.
Fig.~\ref{f0clusters} shows that below $T_c$ indeed three percolating clusters exist -- one for
each of the three center  sectors. Only when one of the sectors starts to dominate at $T_c$,
the subdominant sectors are depleted  and can no longer sustain  percolating clusters. 
Thus the behavior below $T_c$ is similar to the random percolation picture and we conclude that
below $T_c$ the correlations of the phases of the local Polyakov loops at neighboring sites
must be weak.

In the
next section we will analyze the situation of clusters for non-zero cut parameter $f$ which
allows one to focus on the sector that becomes the dominant one above $T_c$.

\begin{figure}[t]
\centering
\hspace*{-5mm}
\includegraphics[width=80mm,clip]{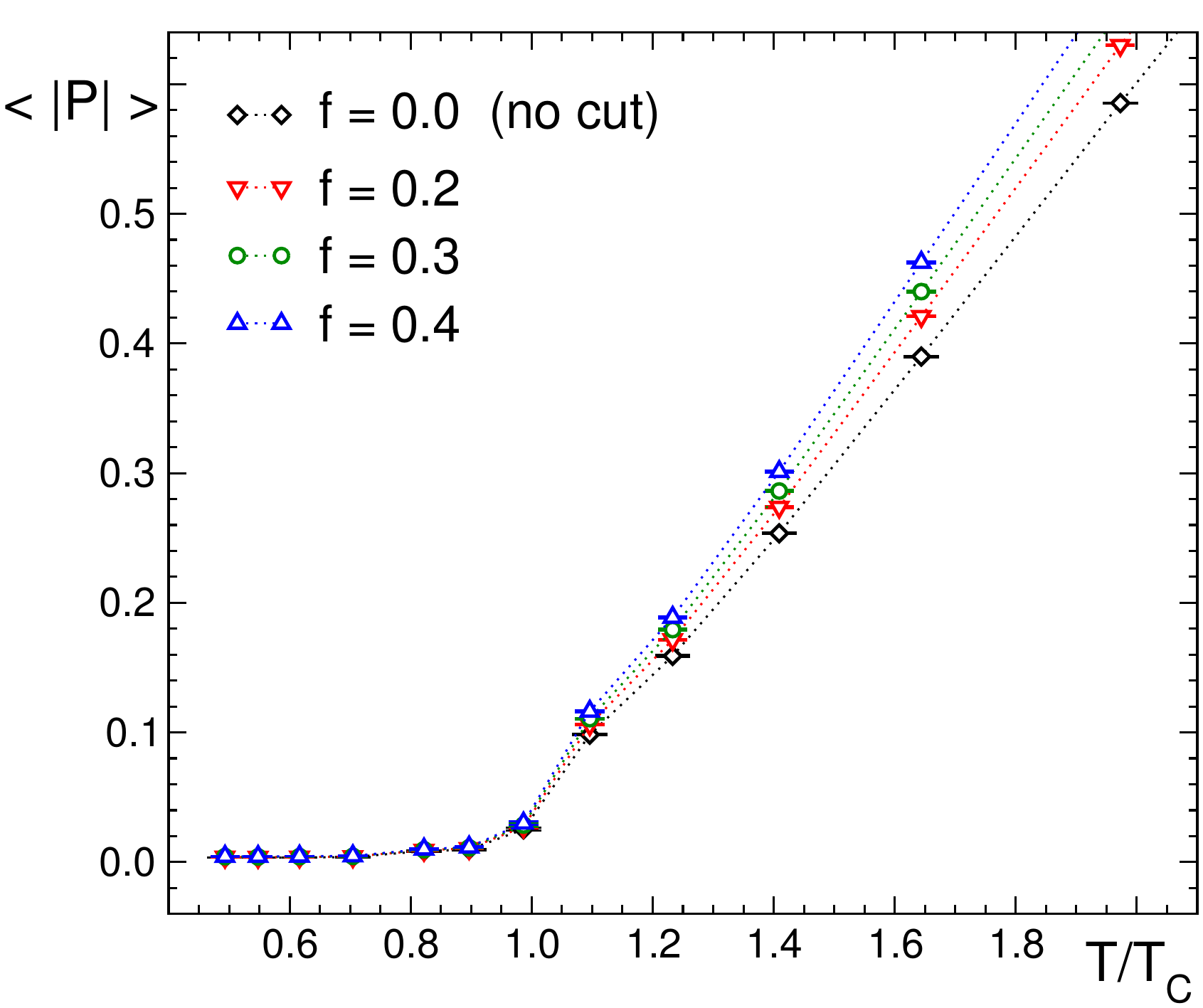}
\caption{Influence of the cut parameter $f$ on the order parameter $\langle |P| \rangle$ 
as a function of the temperature. The results are
for our $\beta = 6.2, 40^3 \times N_t$ ensembles.}
\label{Ploopf_vs_T}
\end{figure}

Before we come to this more detailed analysis of the clusters we inspect how the spatially
averaged  order parameter $\langle |P| \rangle$ reacts to a non-zero value of the cut parameter
$f$. In Fig.~\ref{Ploopf_vs_T} we show $\langle |P| \rangle$ as a function of temperature, but
use only those sites that survive the cut at finite $f$ and compare the results for $f = 0$
(no cut), $f = 0.2, 0.3$ and $f = 0.4$. The plot shows  that the qualitative behavior is
essentially unchanged, in particular near $T_c$. Thus also at finite cut parameter $f$ the
spatially averaged  Polyakov loop $\langle |P| \rangle$ clearly indicates the deconfinement
transition and our procedure for constructing the clusters leaves the fundamental order
parameter intact.

\section{Cluster weight}

\begin{figure}[t]
\centering
\hspace*{-5mm}
\includegraphics[width=80mm,clip]{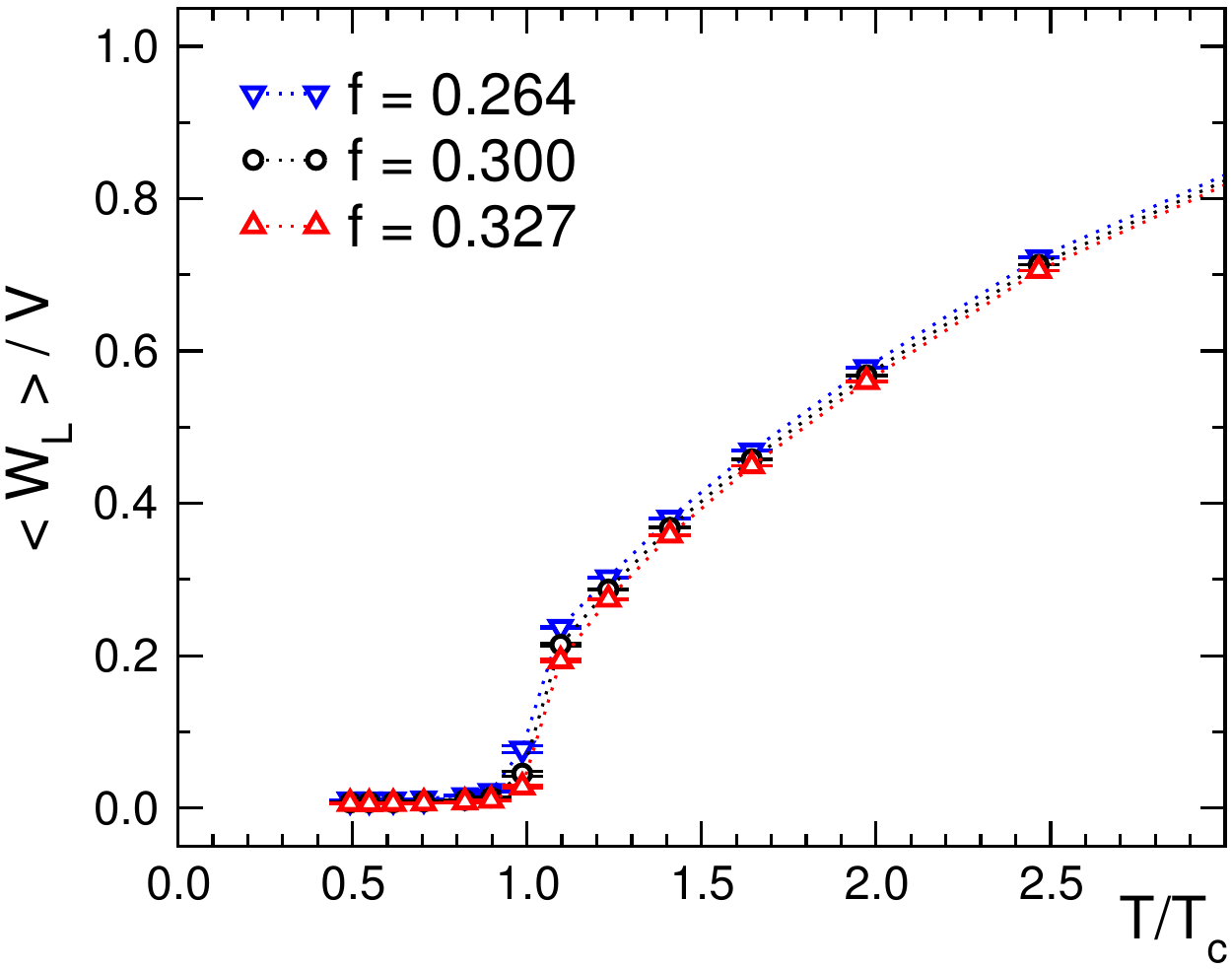}
\caption{Expectation value $\langle W_L \rangle/V$ of the weight of the largest cluster 
normalized by the 3-volume as a function of the temperature. We show results for 
$\beta = 6.2$, $40^3 \times N_t$ and compare 
three different values of the cut parameter $f$.}
\label{weight_fig}
\end{figure}

We begin the analysis of the center clusters by studying the weight $W_L$ of the largest cluster, 
i.e., the number of sites belonging to the largest cluster. In Fig.~\ref{weight_fig} we plot the 
expectation $\langle W_L \rangle / V$ of the weight normalized by the 3-volume $V = N_s^3$ as a function 
of the temperature. We show the results for our
$\beta = 6.2$, $40^3 \times N_t$ lattices, and  compare 
three different values of the cut parameter $f$.  

Fig.~\ref{weight_fig}
shows that there is a window for the cut parameter $f$ where the corresponding
clusters indicate the phase transition: Below $T_c$ the weight of the largest 
cluster is finite and small compared to the volume. At $T_c$ the largest cluster 
starts to grow and occupies a certain fraction of the volume that keeps growing with $T$ towards
the maximum weight $\langle W_{L} \rangle / V = 1$.

\begin{figure}[t]
\centering
\hspace*{-2mm}
\includegraphics[width=80mm,clip]{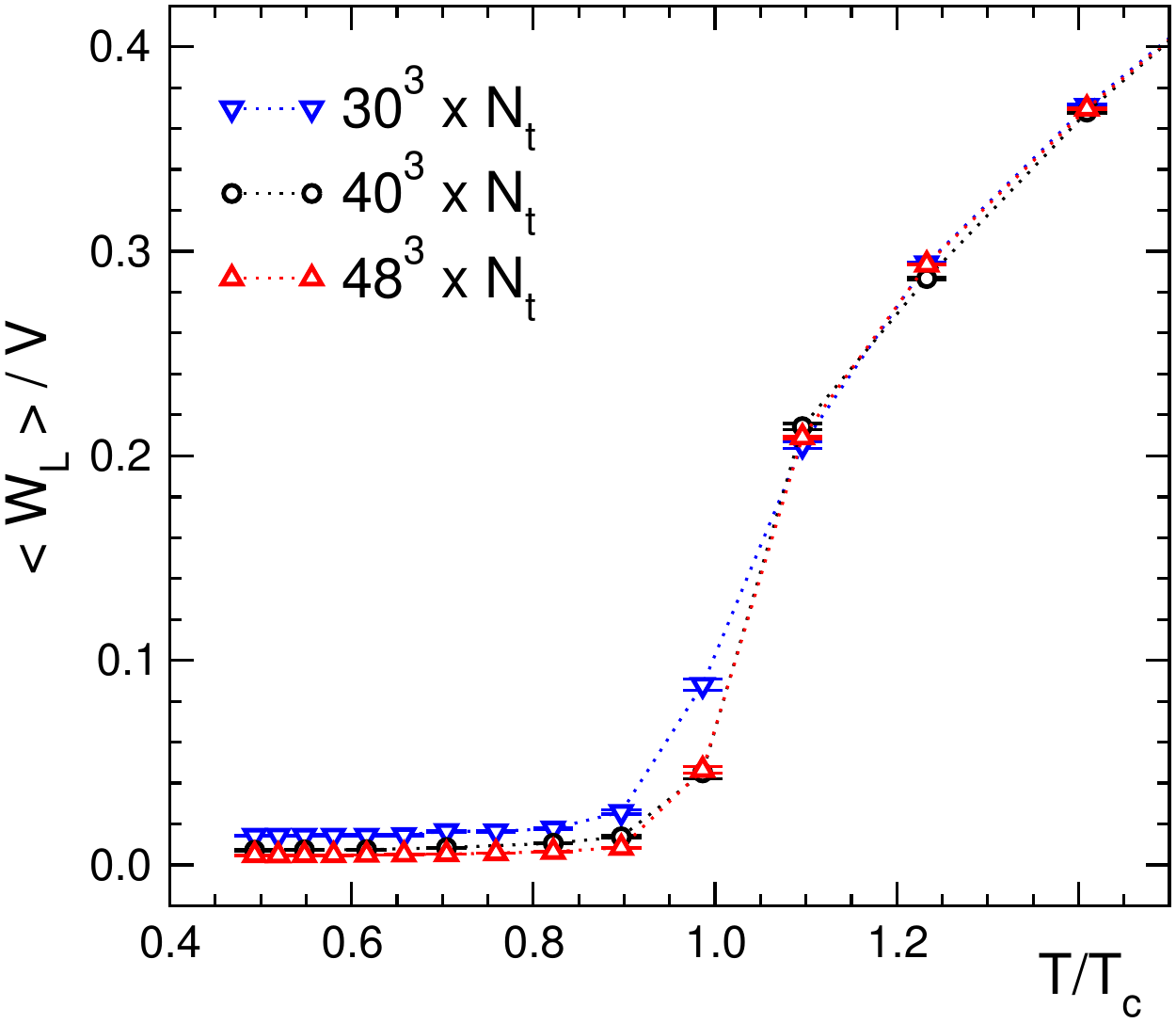}
\caption{Expectation value $\langle W_L \rangle/V$ of the weight of the largest cluster 
normalized by the 3-volume as a function of the temperature. The results are for our 
$\beta = 6.2$ ensembles with a cut parameter of $f = 0.3$ and we compare three different
spatial volumes $V = 30^3$, $V = 40^3$ and $V = 48^3$.}
\label{weight_vol}
\end{figure}

To further understand the behavior of the weight of the largest cluster we repeat the analysis 
of the weight $\langle W_L \rangle / V$ as a function of $T$ for three different spatial volumes
$V = 30^3, 40^3$ and $V = 48^3$. We use the $\beta = 6.2$ ensembles with a cut parameter of $f
= 0.3$ for that study and show the results in Fig.~\ref{weight_vol}. We observe two types of
finite size effects: One is the well known rounding of observables near $T_c$ for smaller volumes. More interesting
is the different behavior below $T_c$. For all three volumes we observe a long plateau below
$T_c$, which is however located at different values of $\langle W_L \rangle / V$. The
interpretation is that for temperatures sufficiently below $T_c$ the largest cluster has essentially a constant 
average weight $\langle W_L \rangle$. When one normalizes to the spatial volume $V$, the
ratio $\langle W_L \rangle/V$ will depend on $V$ and decrease with increasing $V$. For
temperatures sufficiently above $T_c$ the largest cluster extends over all of the volume 
(see below) and fills a fixed fraction of the lattice points. The results for $\langle W_L \rangle / V $ from
different volumes essentially fall on top of each other then.

\begin{figure}[t]
\centering
\hspace*{-5mm}
\includegraphics[width=86mm,clip]{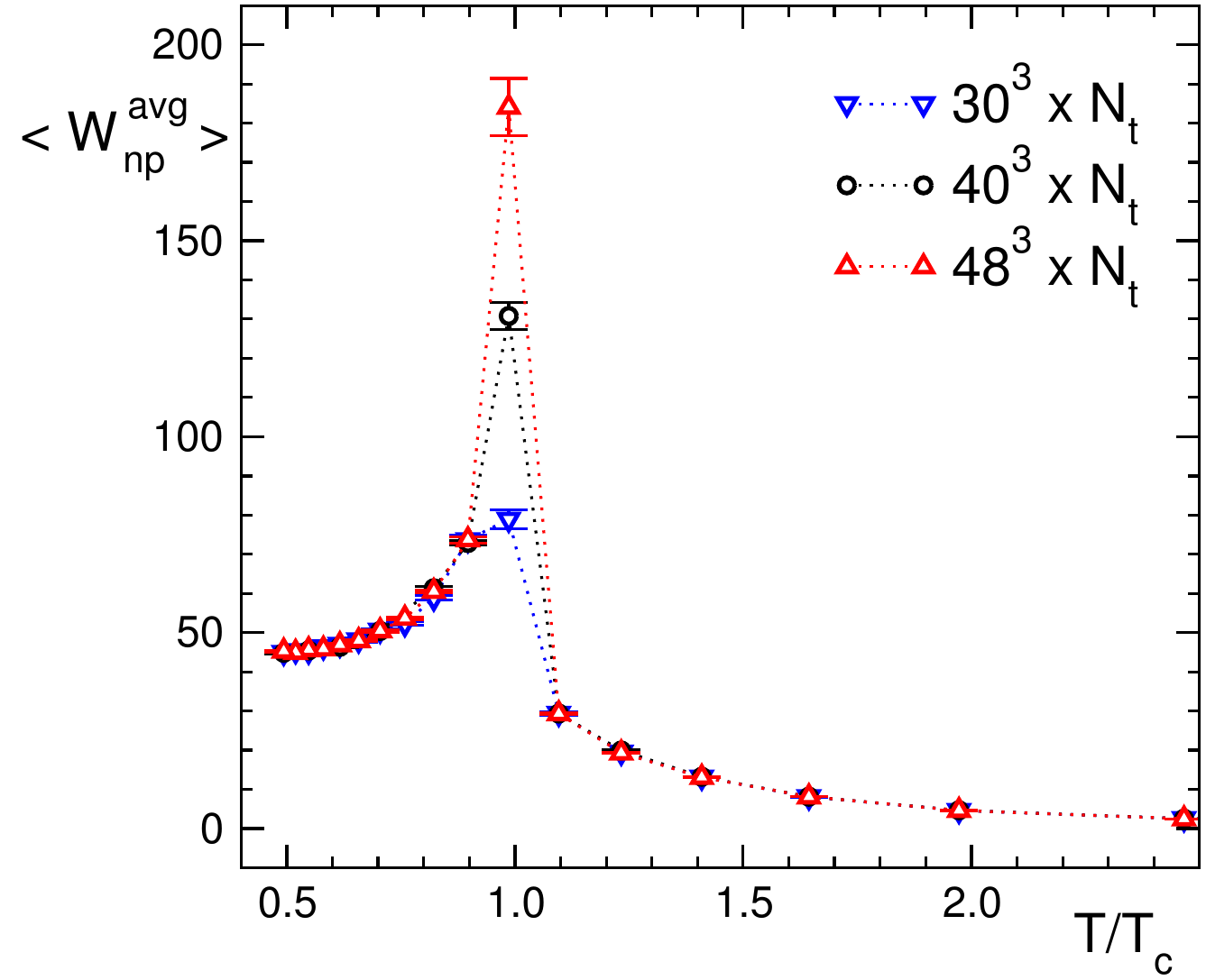}
\caption{Expectation value $\langle W_{\rm np}^{\rm avg} \rangle$ of the average weight 
of non-percolating clusters as a function of the temperature. The results are for our 
$\beta = 6.2$ ensembles with a cut parameter of $f = 0.3$ and we compare three different
spatial volumes $V = 30^3$, $V = 40^3$ and $V = 48^3$.}
\label{weight_np}
\end{figure}

Another important quantity in percolation theory is the average weight of the
non-percolating clusters, which is defined as
\begin{equation}
	W_{\rm np}^{\rm avg} \; = \; \sum_s w_s \, s \; = \; \frac{\sum_s n_s \, s^2}{\sum_{s^\prime} n_{s^\prime} \,  {s^\prime}} \; ,
	\label{eq:npweightdef}
\end{equation}
where the average is taken with respect to 
\begin{equation}
	w_s = \frac{n_s s}{\sum_{s^\prime} n_{s^\prime} {s^\prime}} \; ,
\end{equation}
as is customary in percolation theory~\cite{stauffer1994introduction}. This averaging 
ensures that clusters are represented proportionally to their sizes. 
In the above equations, $s$ is the size of the cluster and $n_s$ is the number of clusters of size $s$. The sums run over all non-percolating clusters. This quantity behaves like a susceptibility and sharply indicates
the position of the transition~\cite{stauffer1994introduction}. In Fig.~\ref{weight_np} we show the expectation value $\langle W_{\rm np}^{\rm avg} \rangle$
for our $\beta = 6.2$ ensembles at $f = 0.3$ and compare all three volumes. The data very cleanly show the peaking behavior of a susceptibility at $T = T_c$,
which becomes more pronounced with increasing volume.
We remark that the largest non-percolating cluster weight, $\langle W_{\rm np} \rangle$, also shows the same behavior as the average defined in Eq. (\ref{eq:npweightdef}) and later we will consider both quantities.

\section{Cluster radius and physical scale}

Let us now proceed by making contact to a physical scale. So far the cut parameter $f$ has been introduced as a parameter in 
our cluster construction. Here we show that it may be related to the size of the clusters at low temperature.  We have seen in the last section
that at fixed $f$ the cluster weight is essentially constant at sufficiently low temperatures.  We now determine the radius of the clusters and study its dependence 
on the cut parameter $f$.  Subsequently we adjust $f$ such that in physical units the radius has some fixed value in the low temperature phase, e.g., 0.5 fm.  This 
can be repeated for all our three couplings $\beta$ and we thus can compare physical results for three different lattice spacings $a$, using the cluster radius in 
fm to set the physical scale.

To describe the linear size of a cluster, we use the center of gravity,
\begin{equation}
x_0 = \sum_{i=1}^s \frac{x_i}{s} \; ,
\end{equation}
to define the cluster radius $R$ as
\begin{equation}
R^2 = \sum_{i=1}^s \frac{|x_i -x_0|^2}{s} \; ,
\label{eq:radiusdef}
\end{equation}
where $x_i$ ($i = 1,2, \ldots, s$) are the sites that belong to a cluster and $s$ is the number of sites in that cluster.
When using periodic boundary conditions, this definition may overestimate the true radius for clusters that are centered across a boundary. 
However, it is then sufficient to consider the set $\{(0,0,0); (N_s/2,0,0); \ldots ;(N_s/2,N_s/2,N_s/2)\}$ of shifted lattice origins, calculate
$x_0$ and $R^2$ using each of them, and finally take the minimal value for the radius. One can show that the unwanted boundary effects indeed 
cancel in this procedure. 

Certainly the radius of an individual cluster for a single configuration is not particularly interesting, but the average over all 
gauge configurations is a meaningful observable, and in the following discussion we always refer to the gauge average when we talk 
about the radius of clusters. This will either be the radius of the largest cluster or that of the largest non-percolating cluster.

Properties of the clusters, and in particular also their radii, will depend on the cut parameter $f$. Thus, from now on, we display $f$ as an argument when considering 
the radius $R(f)$.  The radius of the clusters defined in Eq.~(\ref{eq:radiusdef}) is given in lattice units, 
therefore the radius in fm is obtained after multiplication by the lattice constant
$a$ in fm,
\begin{equation}
 R_{\rm phys} \; = \; a \cdot R(f) \; .
\label{physradius}
\end{equation}
Clearly, at different values of the lattice spacing $a$, the same physical radius will be realized by 
different $R(f)$ and thus by different values of $f$. 

For a comparison of the results from lattices with different $a$, suitable observables in physical units have to be matched. 
As already announced, here we use the physical radius $R_{\rm phys}$ of the largest clusters at a fixed (low) temperature $T<T_c$, where all clusters are finite. In other 
words we fix $R_{\rm phys}$ to some value we want to study, e.g., $R_{\rm phys} = 0.5$ fm, and identify the value of $f(a,R_{\rm phys})$ that, for the 
lattice spacing $a$ we work at, gives rise to $R_{\rm phys} = 0.5$ fm. Using the value $f(a,R_{\rm phys})$ in the cluster construction we can compare
in a meaningful way the properties of the clusters for ensembles at different $a$.

\begin{figure}[t!]
 \centering
 \hspace*{-6mm}
 \includegraphics[width=8cm]{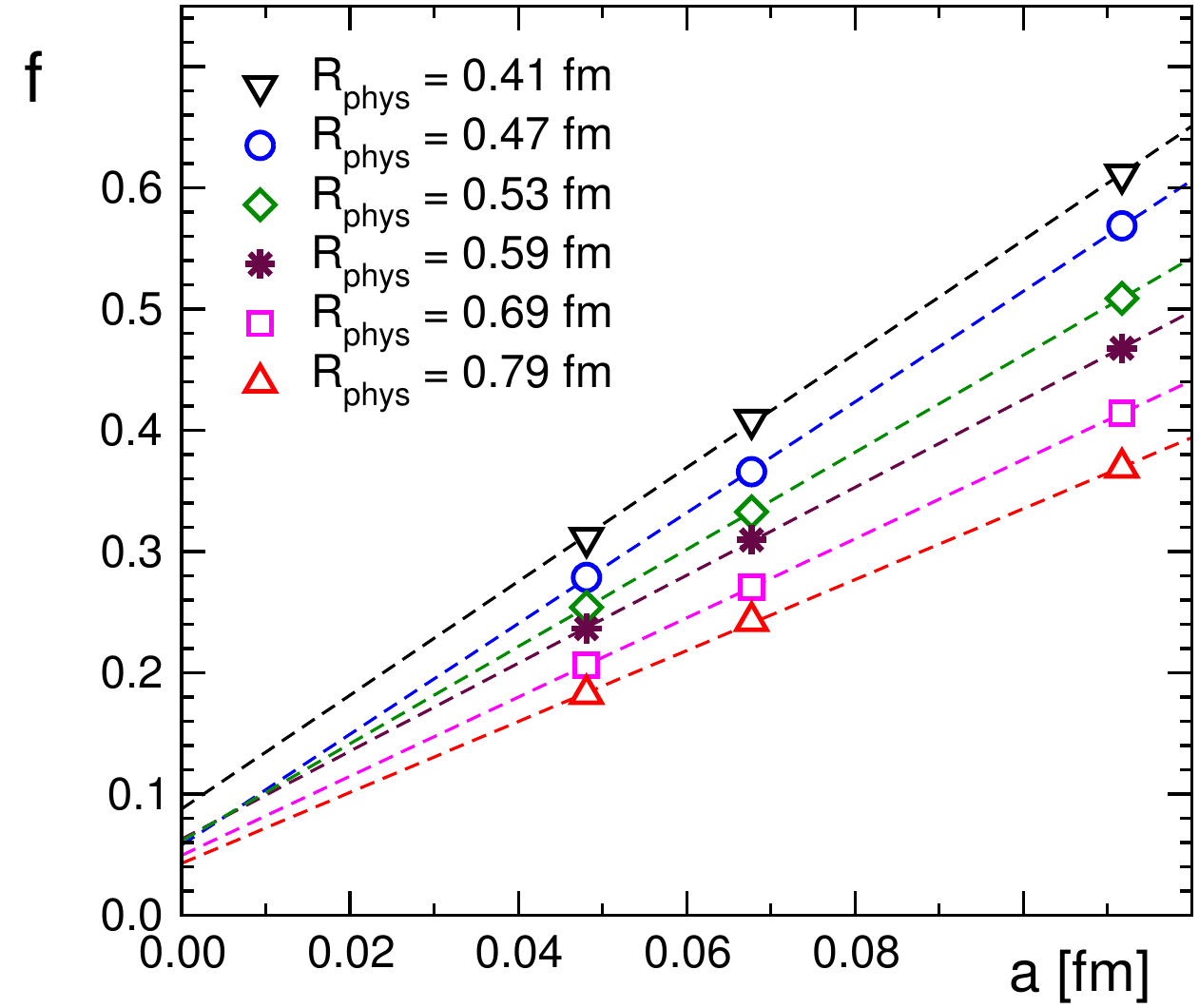}

\vskip5mm 
 
 \hspace*{-6mm}
 \includegraphics[width=8cm]{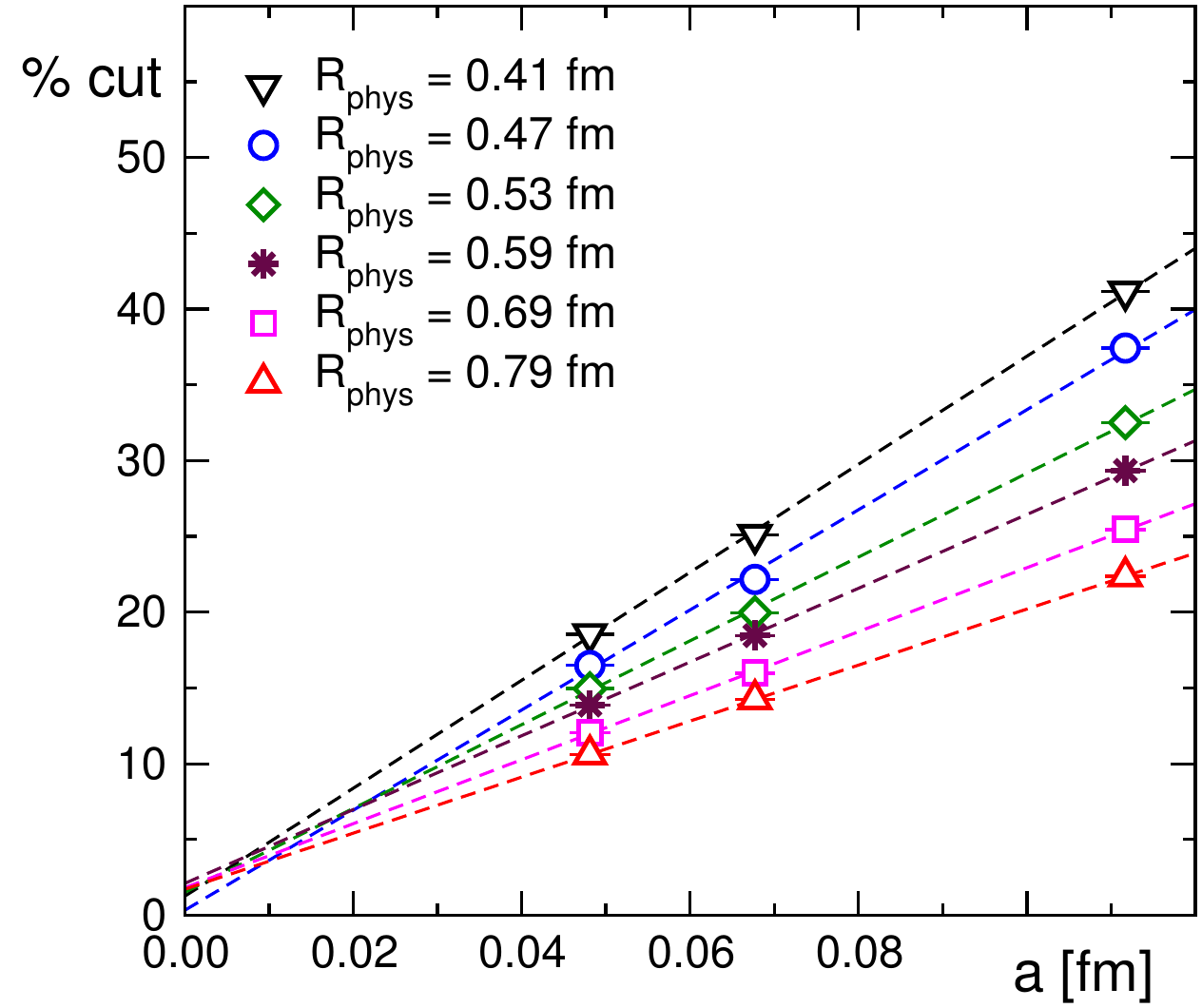}
 \caption{\label{fig:fa}Dependence of the cut on the lattice constant $a$. In the top panel we show the cut parameter $f(a,R_{\rm phys})$ that is 
 necessary to obtain a given $R_{\rm phys}$ of the largest clusters for several values of $R_{\rm phys}$. In the bottom plot we show the corresponding percentage of sites that 
 has to be cut. Both plots are for our $48^3 \times 20$ lattices, i.e., the lowest temperatures we consider.}
 \end{figure}

This procedure defines the ``lines of constant physical
radii'' as a function of $a$. To study these, in the top panel of Fig.~\ref{fig:fa} we show $f(a,R_{\rm phys})$  as a function of the lattice spacing $a$ for different 
values of $R_{\rm phys}$, ranging from 
$0.4$ fm to $0.8$ fm (always using the lowest available temperature, i.e., the $48^3 \times 20$ lattices to determine $f(a,R_{\rm phys})$).

An alternative way of presenting the information about the lines of constant physics is to look at the number of 
sites that are removed by the cut. In the lower panel of Fig.~\ref{fig:fa} we plot the percentage of sites that were cut, again as a function of $a$ for different 
values of  $R_{\rm phys}$. Applying linear fits we find that all data sets extrapolate to a small value between 1\% and 2\%. 

Let us try to interpret this outcome 
in the context of random percolation (which is of course not the full picture as we have correlations here): 
For random percolation, the critical probability is $31.2\%$~\cite{stauffer1994introduction,critperc}. This means that if the number of sites that survive the cut is larger than $3 \times 31.2 \% = 93.6 \%$ percent,
we can have percolating clusters in all three center sectors. The lower panel shows that in the $a \rightarrow 0$ limit the number of sites that are removed is not more than  2 \% and indeed 
sufficiently many sites remain such that in the limit $a \rightarrow 0$ percolating clusters appear in all sectors. This implies that in lattice units $R(f)$ diverges, which is a necessary prerequisite 
that the physical radius  $R_{\rm phys} \; = \; a \cdot R(f)$ remains finite in the continuum limit $a \rightarrow 0$. We stress again, that this picture is based on random percolation and in 
the case of SU(3) gauge theory receives corrections from correlations. However, we will argue below that these are small.

\begin{figure}[t!]
 \centering
 \hspace*{-8mm}
 \includegraphics[width=8.4cm]{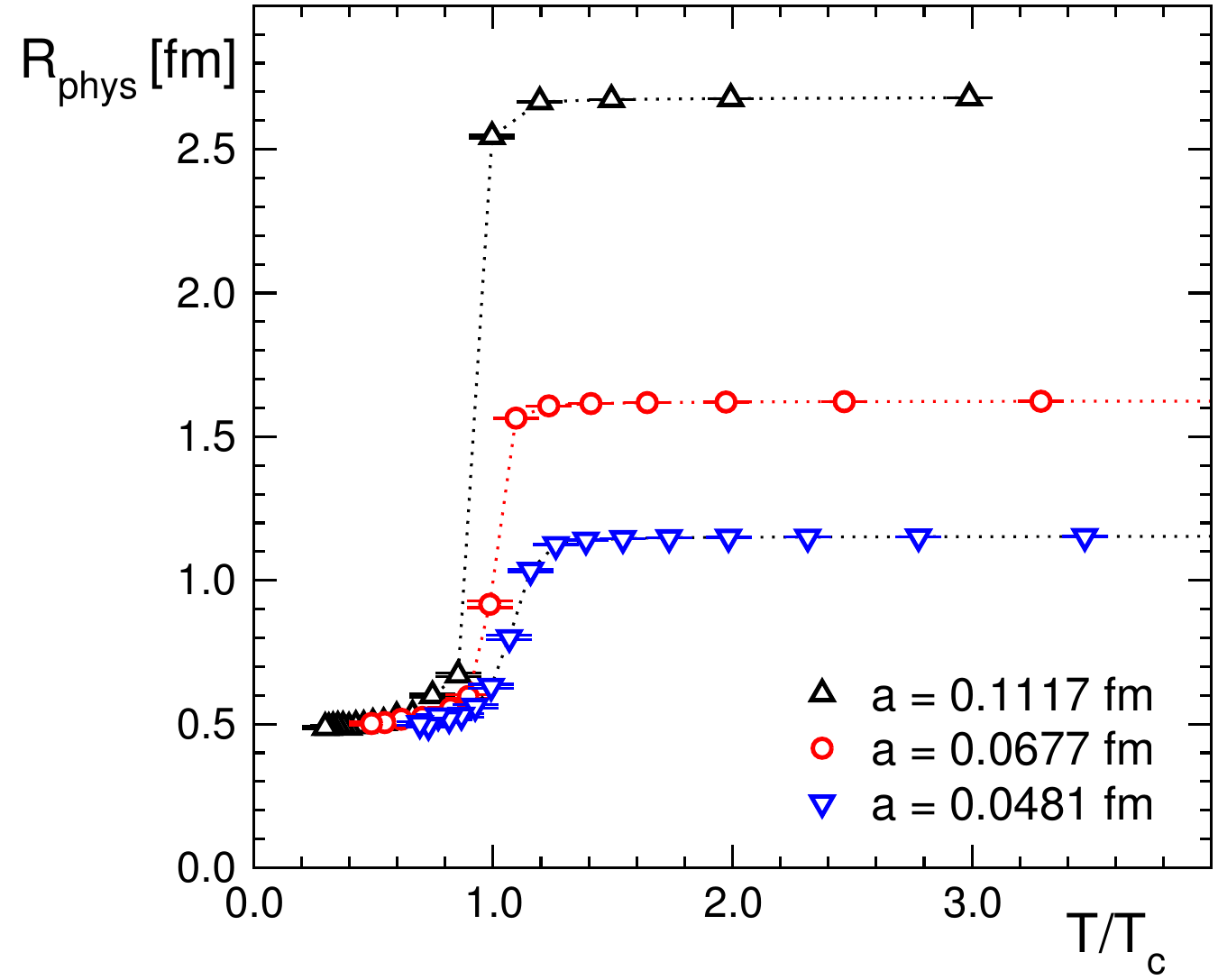}
 \caption{\label{fig:radius}Radius of the largest cluster as a function of the temperature. We compare the results from our 
 $48^3 \times N_t$ lattices for all three values of the lattice constant $a$. }
\end{figure}

Now that we have set the scale at $T<T_c$, we can discuss the continuum 
limit of observables that are sensitive to the transition to the deconfined phase. 
For such a comparison from now on we will use  $R_{\rm phys}=0.5 \textmd{ fm}$ 
for the largest clusters
set at the lowest available temperature. The corresponding values of 
$f(a,R_{\rm phys})$ are given by  $f(a,R_{\rm phys}) =0.543$, $0.355$ and $0.268$ 
for our $a = 0.1117$, $a = 0.0677$ and $a = 0.0481$ lattices (corresponding to $\beta=5.9$, $6.2$ and $6.45$).  
In all subsequent plots the scale was set according to this prescription. 

Before we come to discussing other observables in subsequent sections, 
in Fig.~\ref{fig:radius} we show the physical radius of the largest cluster as a function of the temperature.
We compare results from our $48^3 \times N_t$ ensembles for all three values of the lattice spacing.
Below $T_c$ the radius remains close to the radius of $R_{\rm phys}=0.5 \textmd{ fm}$ which was used 
to set the scale. Near $T_c$ we see a rapid increase of the
radius, which corresponds to the fact that center symmetry is broken and a single large cluster spreads 
over all of the lattice. Above $T_c$ the radius saturates 
at a value which is half of the spatial extent of the lattice in physical units.

\section{Percolation}

\begin{figure}[t]
\centering
\hspace*{-4mm}
\includegraphics[width=80mm,clip]{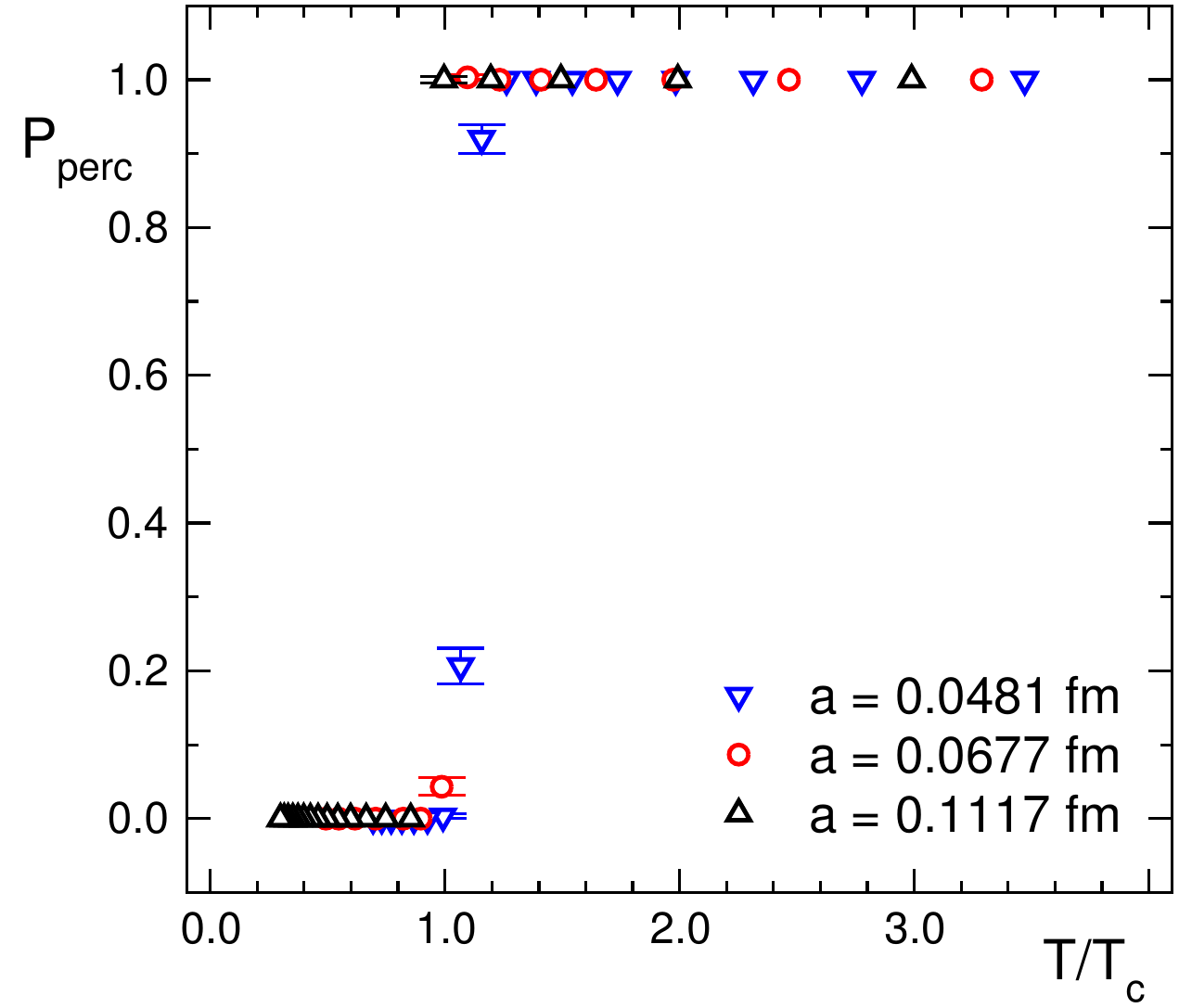}
\caption{The probability $P_{\rm perc}$ of finding a percolating cluster as a function of the 
temperature. We use our $48^3 \times N_t$ ensembles and show results for all three values of the lattice 
spacing $a$.
\label{percprob}}
\end{figure}

An interesting question is whether the maximal clusters start to percolate at $T_c$.
Percolation is analyzed in Fig.~\ref{percprob} where we show the probability $P_{\rm perc}$
of finding a percolating cluster as a function of $T$, where  $P_{\rm perc}$ is again the average over all gauge configurations.  
We use $48^3 \times N_t$ lattices  and 
compare the results for the three different values of the lattice constant $a$. 
The scale was set as discussed in the previous section. We observe that the probability is very 
close to 0 for all temperatures $T$ below $T_c$, with a step-like increase to 1 for 
$T > T_c$. For the finest lattice (which is also the smallest physical volume here) there is a visible rounding 
of the step-like behavior, which we attribute to finite size effects.

\section{Fractal dimension}

\begin{figure}[t!]
\centering
\hspace*{-6mm}
\includegraphics[width=85mm,clip]{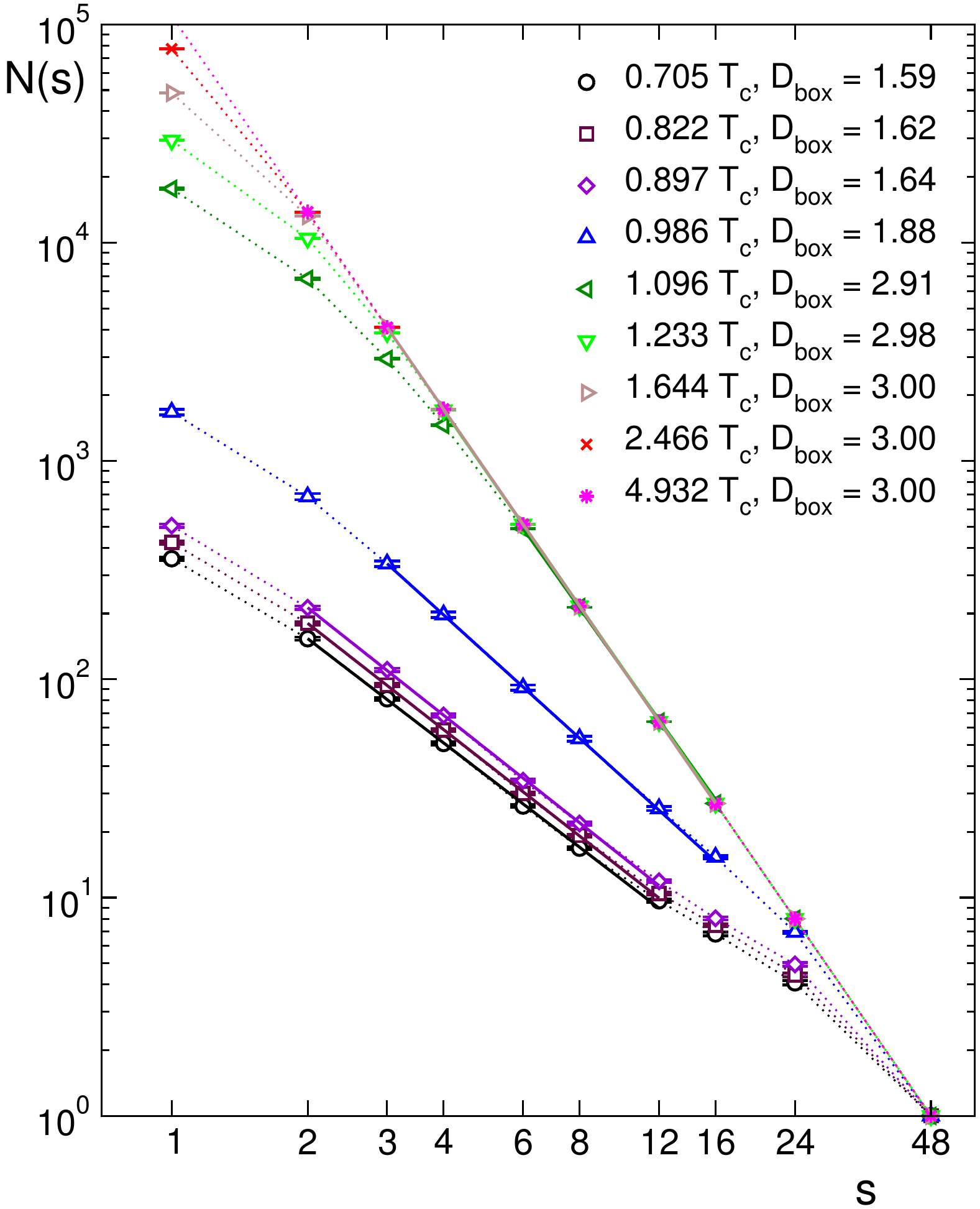}
\caption{Example for the determination of the box-counting dimension of the largest cluster ($48^3 \times N_t$, $\beta = 6.2$). We show the data (symbols connected with dotted lines) 
for $N(s)$ versus temperature, both
on logarithmic scales. In the range of $s$ values we used for the fit we overlay with full lines the fit function $C \, s^{-D_{\rm box}}$, which on the log-log scale appears as a straight line. The corresponding 
fit value for the parameter $D_{\rm box}$ is also given in the legend. }
\label{boxcount_fig}
\end{figure}

We continue our analysis of the clusters by analyzing the fractal dimension of the largest cluster. One definition we use is the box counting dimension where 
one counts the number $N(s)$ of cubes of linear size $s$ that are needed to cover the whole cluster. 
In the limit of small $s$ one expects the behavior $N(s) \sim s^{-D_{\rm box}}$, with the exponent $D_{\rm box}$ 
defining the fractal dimension. 

Fig.~\ref{boxcount_fig} illustrates the approach for the determination of the 
box counting dimension of the largest cluster ($48^3 \times N_t$, $\beta = 6.2$). 
For $N_s = 48$ we use cube sizes $s=$ 1,2,3,4,8,12,16,24, and 48 i.e., all divisors of 48.
In Fig.~\ref{boxcount_fig} we show $N(s)$ versus $s$ in a log-log plot comparing the results for several temperatures. The fractal dimension $D_{\rm box}$ is found by fitting the data to the function $N(s) = C s^{-D_{\rm box}}$.
In the figure we connect the data points used for the fit by full lines and in the legend quote the fit results for $D_{\rm box}$. 
Large and small values of $s$ were omitted in the 
fit since they suffer from finite volume effects (for $s$ close to 48) 
and the fact that there is a smallest scale which introduces an ultraviolet cutoff.

\begin{figure}[t!]
\centering
\hspace*{-4.5mm}
\includegraphics[width=75mm,clip]{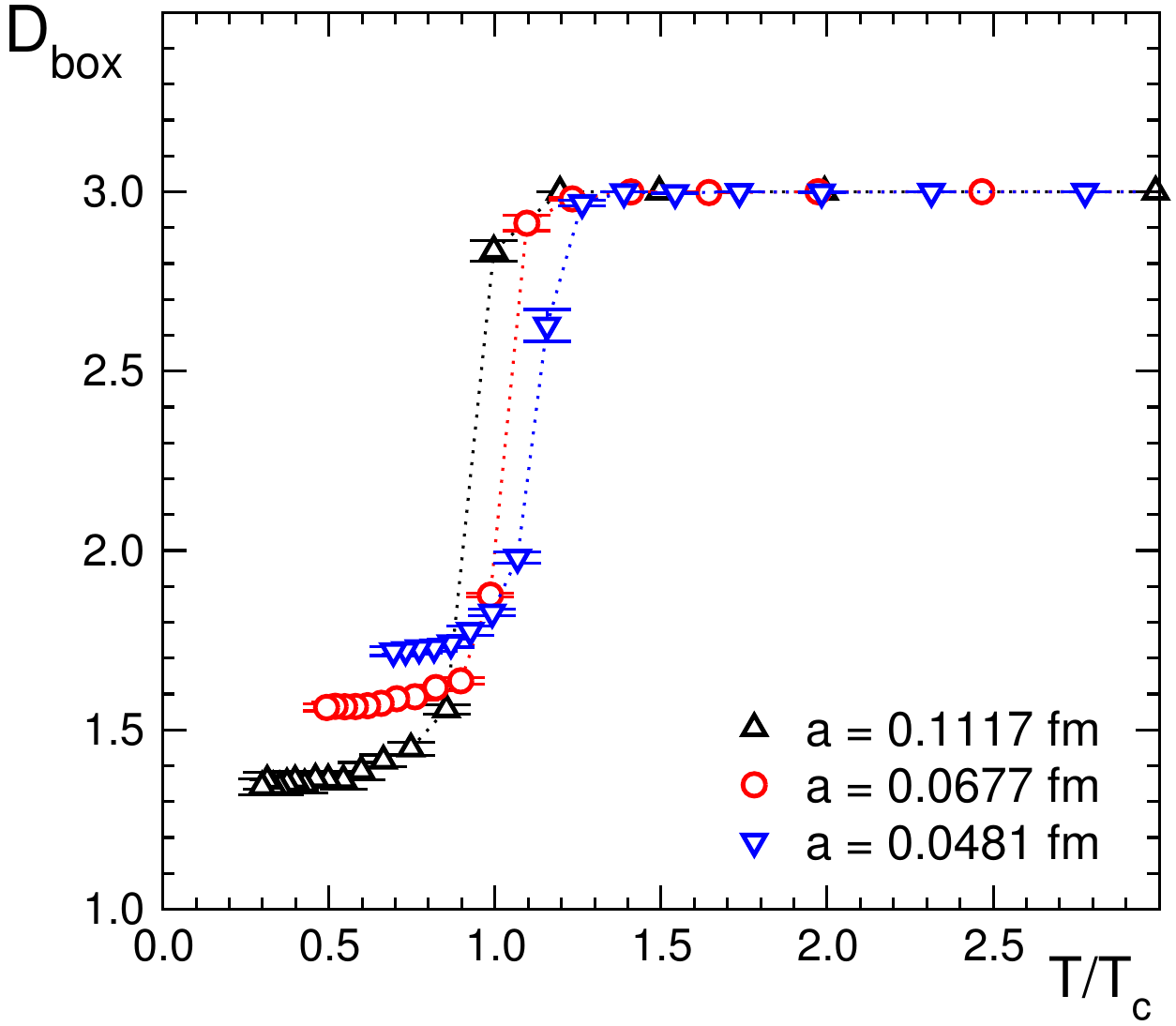}
\caption{The fractal dimension from the box counting method as a function of the temperature. 
We show results for our $48^3 \times N_t$ lattices for all three lattice spacings.}
\label{Dbox_vs_T}
 
 \vskip5mm
 
 \centering
  \hspace*{-3mm}
 \includegraphics[width=75mm]{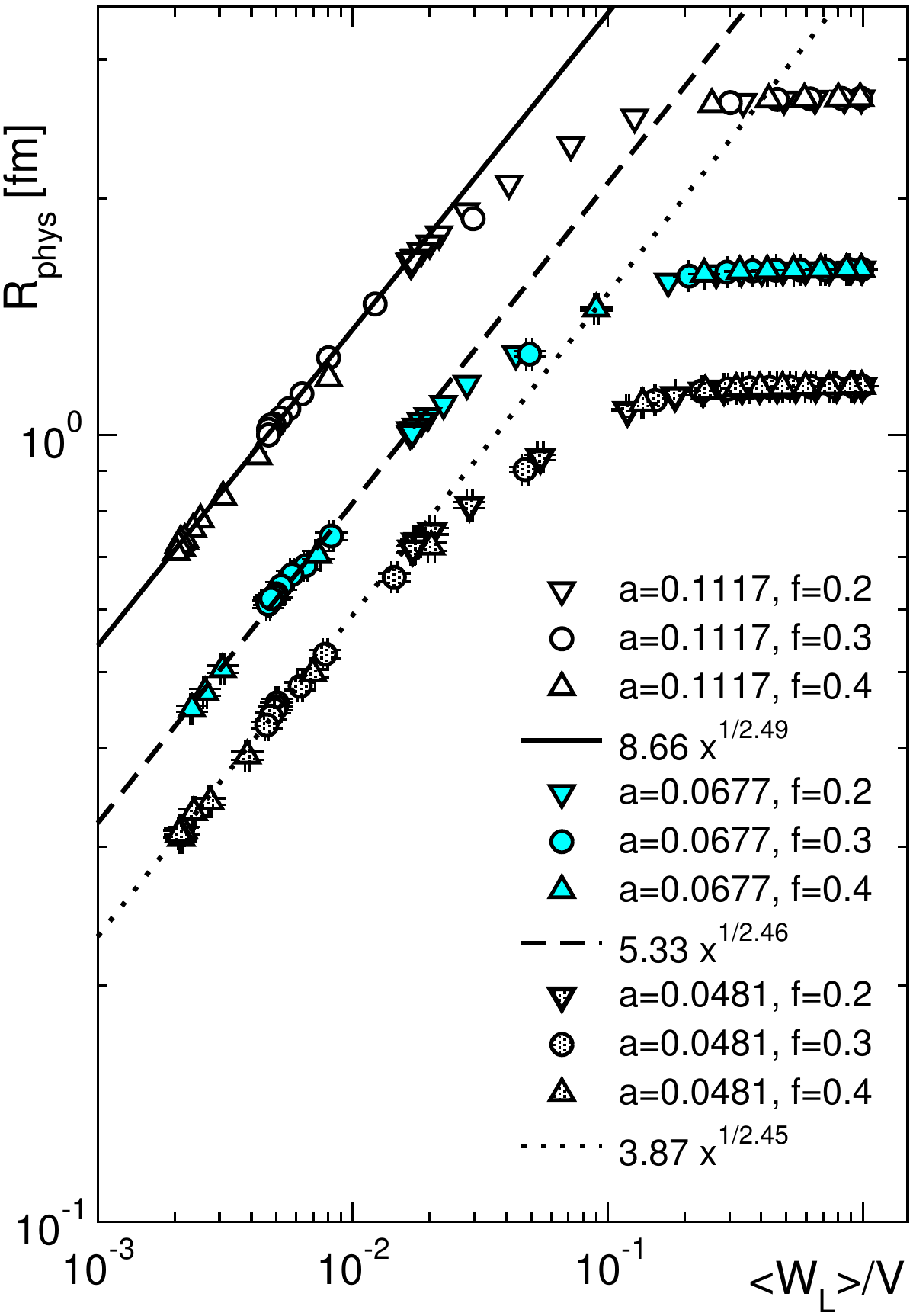}
 \caption{\label{fig:universal}Radius of the largest cluster against its weight. We show the results for different values of $f$ and compare all three lattice spacings on our $48^3 \times N_t$ lattices. In the plot we also show the results of fits of the asymptotic behavior at small 
 $\langle W_L \rangle$ with a power law $R_{\rm phys} \propto \langle W_L \rangle ^{1/D_{\rm sc}}$.}
\end{figure}

The figure shows that away from the smallest and largest values of $s$ we indeed 
observe a linear behavior which can be fit reliably. Also it is obvious that the slope becomes steeper with 
increasing temperature indicating that the dimension increases with $T$. 

In Fig.~\ref{Dbox_vs_T} we show 
the fractal dimension from the box counting method as a function of the temperature 
for our $48^3 \times N_t$ lattices with three different lattice spacings. The plot nicely illustrates 
that the box counting dimension is an increasing function of $T$, starting  near 
$D_{\rm box} \sim 1.4-1.7$ below $T_c$ and reaching $D_{\rm box} = 3$ in a rather rapid increase near $T_c$.
Besides these qualitative characteristics, we also observe that lattice discretization 
effects become enhanced in the confined phase, showing a slow scaling of $D_{\rm box}$ towards 
the continuum limit. 

Based on the above results from the box counting method, we conclude that while the largest clusters are 
three-dimensional objects in the deconfined phase, they are fractals below $T_c$ with a dimension significantly 
smaller than 3. 

\begin{figure}[t!]
 \centering
 \hspace*{-1mm}
 \includegraphics[width=73mm]{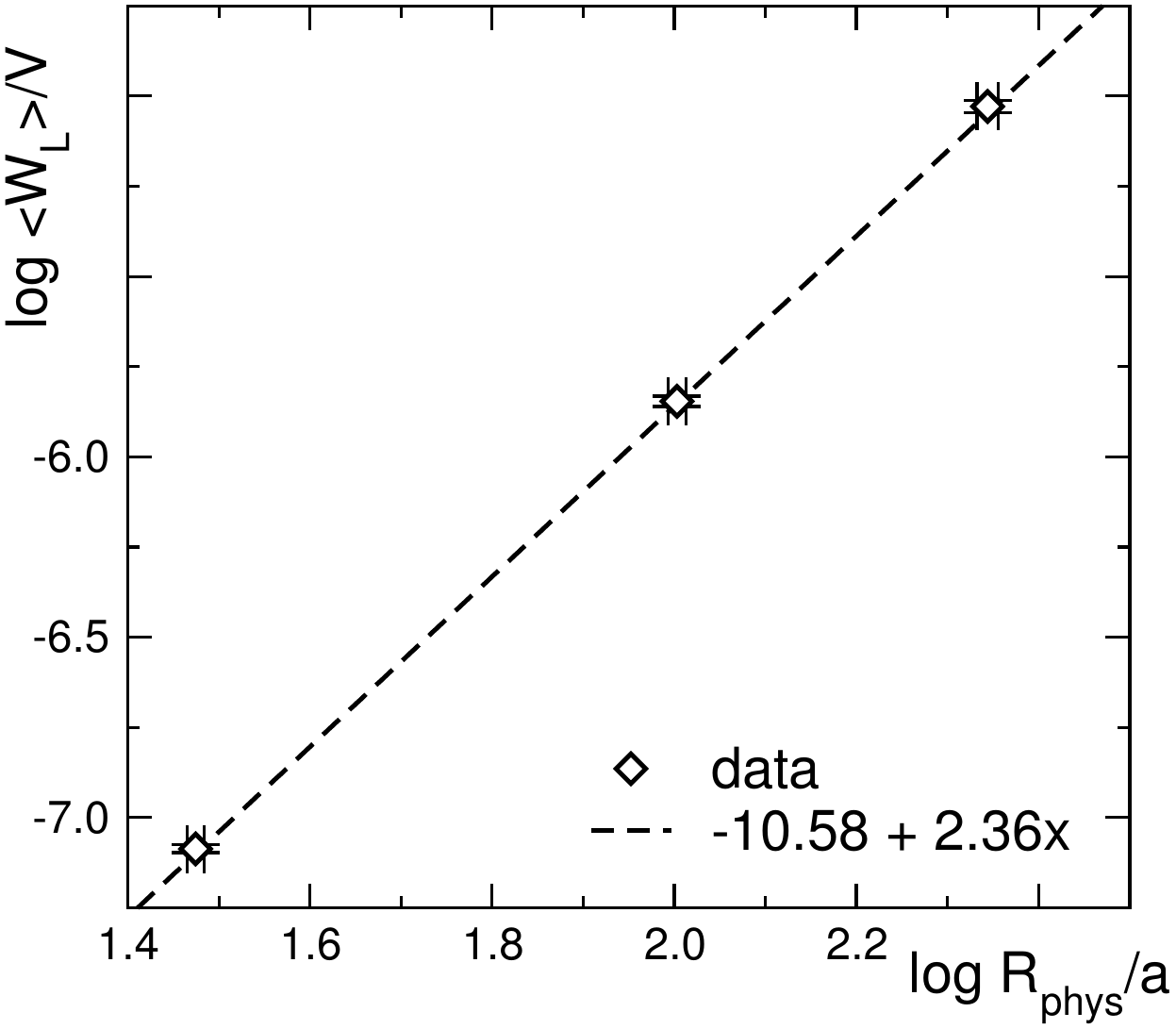}
 \caption{\label{fig:fracdim_sR} Logarithm of the weight of the largest cluster on our $48^3 \times 20$ lattices versus the 
 logarithm of its radius for the three different lattice spacings. The data (symbols) are fit with a straight line (dashed). }
\vskip5mm
 \centering
  \hspace*{-4mm}
 \includegraphics[width=73mm]{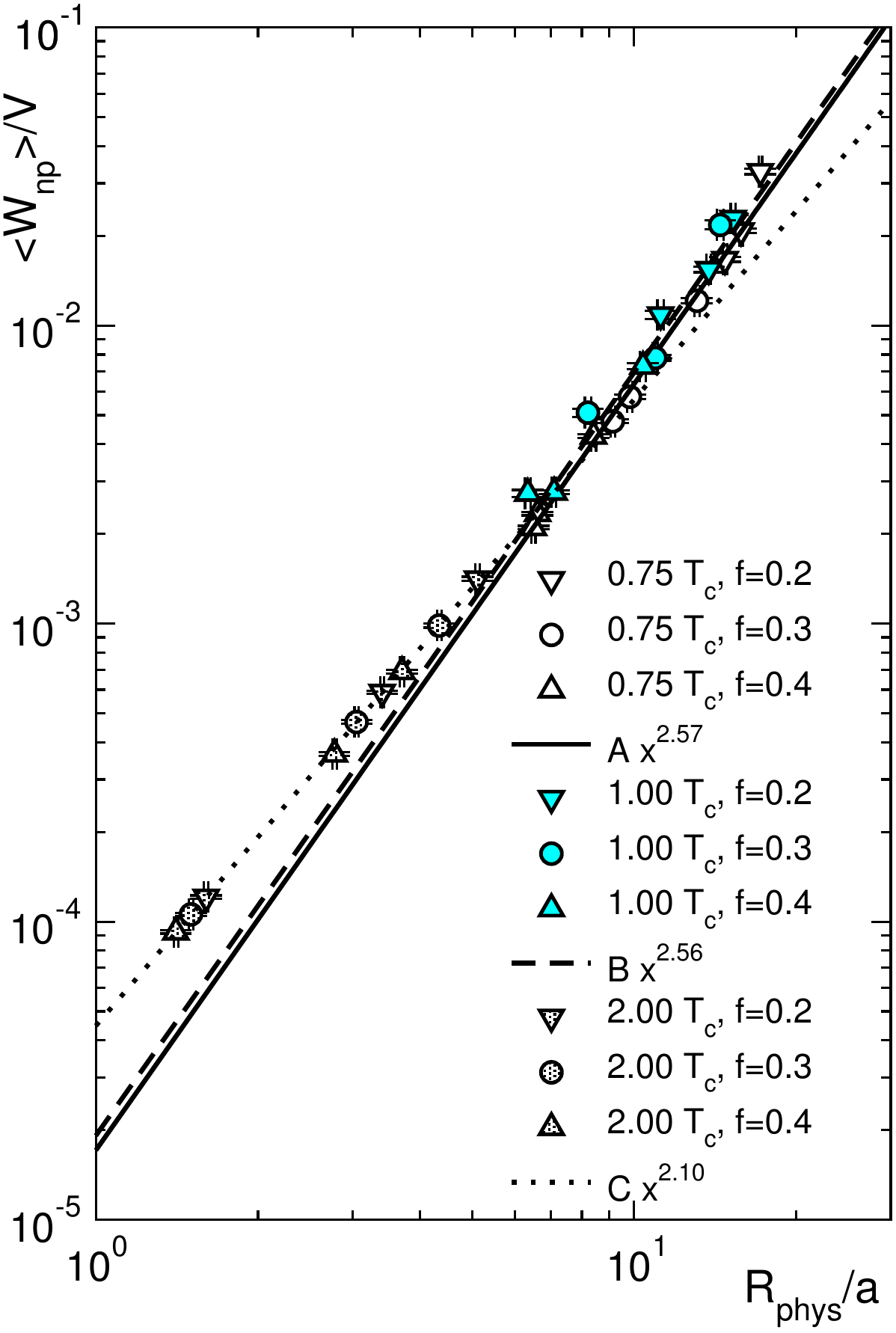}
 \caption{\label{fig:universal_np}Weight of the largest non-percolating cluster against its radius for three representative values of the temperature. Results are shown for 
 various values of $f$ and the lattice spacing, as measured on our $48^3 \times N_t$ lattices. 
 The lines indicate fits for the scaling dimension as described in the text.}
\end{figure}

It is well known that different definitions of the fractal dimension may give slightly different results, see, e.g., Ref.~\cite{fractaldim}.
To further support the fractal nature of the clusters at low temperatures, we turn to another approach 
to determine their dimensionality, by considering how their weight scales with their radius in lattice units. 
This `scaling' dimension $D_{\rm sc}$ is defined by the relation
\begin{equation}
\langle W \rangle \; \propto \; R^{D_{\rm sc}} \; , 
\label{eq:Dscdef}
\end{equation}
i.e., the response of the average weight $\langle W \rangle$ to a change of the radius $R$. 
We will consider Eq.~(\ref{eq:Dscdef}) to define the dimensionality of both the largest and the largest non-percolating cluster.
In our setting $\langle W \rangle$ and $R$ depend both on the cut parameter $f$ and on the lattice constant $a$. Thus, either of them 
can be used to change $\langle W \rangle$ and $R$ so that the relation in Eq.~(\ref{eq:Dscdef}) can be studied.

First, we consider the cut parameter $f$ as the driving parameter: Fig.~\ref{fig:universal} 
shows the relation between the average weight $\langle W_L \rangle$ of the largest cluster and the radius $R_{\rm phys}$ for various values of $f$ and of $a$.
On the log-log scale the data show a linear behavior for small $\langle W_L \rangle$, which corresponds to small $T$. We fit this small $T$ 
behavior 
according to $R  \propto \langle W_L \rangle^{1/D_{\rm sc}}$ and find $D_{\rm sc} =2.49(4)$, $2.46(4)$ and $2.45(5)$ for the three lattice spacings (see the figure). 
This spread of the values for $D_{\rm sc}$ from the different lattice constants is smaller than that for $D_{\rm box}$ ($D_{\rm box} \sim$ 1.4 -- 1.7), 
indicating that the discretization errors in $D_{\rm sc}$ are smaller than those in $D_{\rm box}$.

Second, we may also use the lattice spacing to drive $\langle W_L \rangle$ and $R$. In Fig.~\ref{fig:fracdim_sR}, the logarithm 
of the average weight $\langle W_L \rangle$ of the largest clusters is plotted as a function of $\log R$ for the lowest temperature for our three lattice spacings. The data points from 
the three lattice spacings fall on a straight line, and a fit according to  $\langle W_L \rangle \propto R^{D_{\rm sc}}$  gives $D_{\rm sc}=2.36(2)$, a value in the vicinity 
of the result obtained when varying $f$.

To complete the discussion of the fractal dimension, in Fig.~\ref{fig:universal_np} we 
perform a similar analysis for the non-percolating clusters.
This time the weight $\langle W_{\rm np} \rangle$ of the largest non-percolating cluster 
is plotted against its radius for three different temperatures. 
We find that on the log-log scale the data 
essentially fall on straight lines for each temperature, irrespective of whether 
we vary the lattice spacing or the cut parameter. This allows for 
a simultaneous fit of the data at various $a$ and $f$ for each temperature, as 
depicted in the figure. 
Note that, contrary to the case of the largest clusters, here the linear behavior 
persists also for high temperatures, since the non-percolating clusters are always finite and, thus, do not suffer from finite size effects. 
A fit of the linear behavior with $\langle W_{\rm np}\rangle \propto R^{D_{\rm sc,np}}$ gives 
$D_{\rm sc,np} = 2.57(8)$, $2.56(15)$ and $2.10(1)$ for the three temperatures $T/T_c=0.75$, $1$ and $2$, respectively. 
This 
indicates that the non-percolating clusters keep a fractal dimension  $D_{\rm sc,np} < 3$ also across the deconfinement transition. 

The analysis of the fractal dimension can be summarized as follows: Below $T_c$ all clusters have fractal dimensions considerably below 3 ($D_{\rm box} \sim$ 1.4 -- 1.7, and $D_{\rm sc}=D_{\rm sc,np} \sim$ 2.5 -- 2.6 for the largest (non-percolating) cluster). At $T_c$ a percolating cluster emerges with a dimensionality that quickly reaches $3$, while the non-percolating clusters remain fractal ($D_{\rm sc,np}\sim 2.1$) across the transition.
We remark that the fractal dimension of clusters in random percolation theory close to the critical occupation probability equals 
$2.52$~\cite{quenched1}, a value quite close to our findings for $D_{\rm sc}$, again indicating that correlations between neighboring Polyakov loop phases are small.

\section{Surfaces of clusters}
\label{sec:surfaces}

Let us finally come to a question that was part of the initial motivation for this study: The possibility that different center domains are separated by domain walls that play a role in the evolution and phenomenology of the 
Quark Gluon Plasma in heavy-ion collisions~\cite{centermodels}. From the analysis of the fractal dimension of the clusters discussed in the previous section it is already clear that the interfaces between 
center domains will not be smooth two-dimensional walls. Still it is interesting to study how the surface depends on the temperature and also the relation between surface and weight of the clusters 
provides information about the type of percolation that dominates the behavior near $T_c$.

\begin{figure}[t]
 \centering
 \hspace*{-6mm}
 \includegraphics[width=75mm]{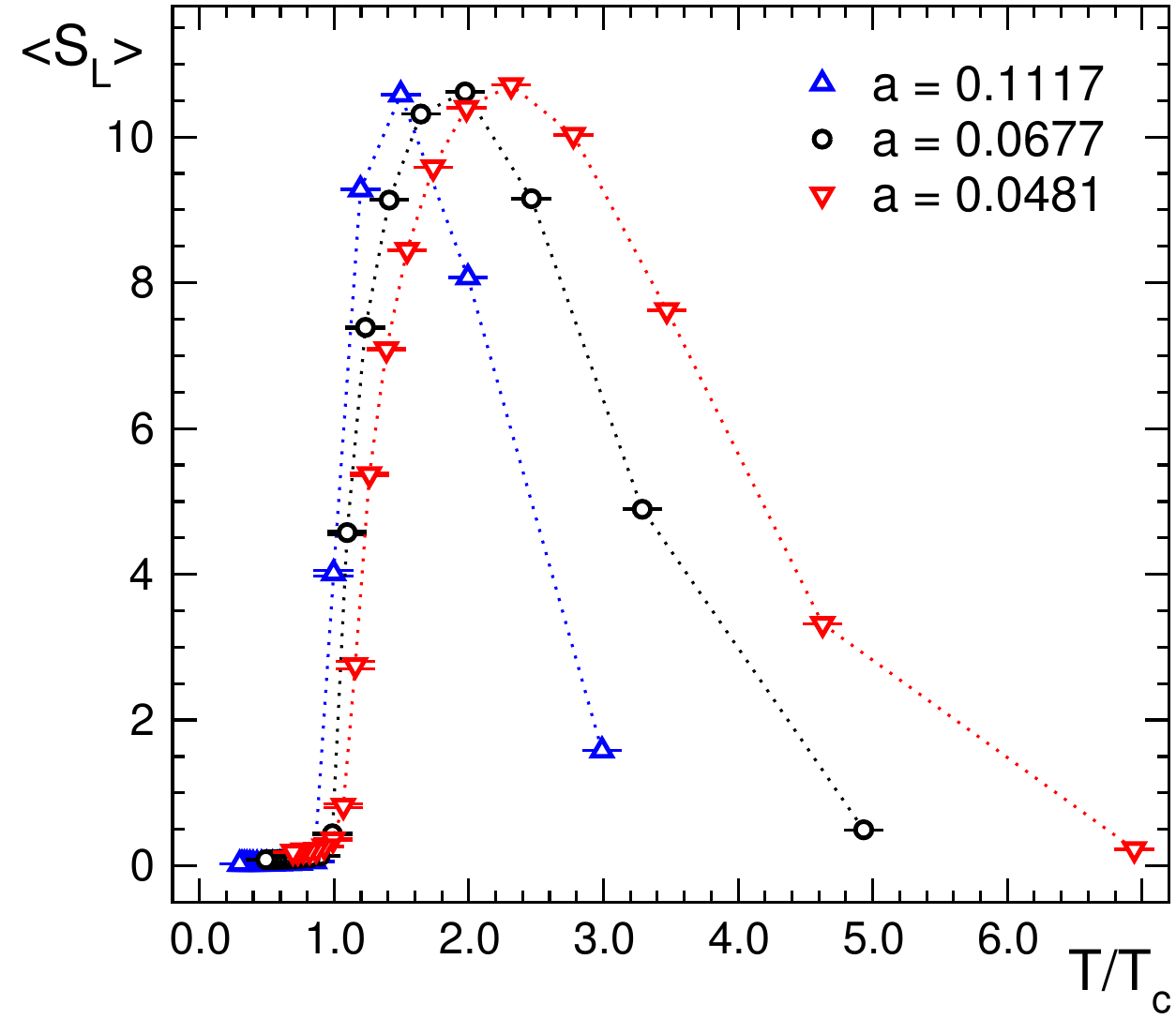}
 \caption{\label{fig:surface_vs_T} Average surface $\langle S_L \rangle$ 
 of the largest cluster as function of the temperature. Results are shown for our $48^3 \times N_t$ lattices, for all three values of the lattice constant. }
\end{figure}

We define the surface $S$ of a cluster as the number of links where one site of the link is a member of the cluster, while the other is not. This is equal to the number of plaquettes on the dual lattice
needed to wrap the cluster. $S$ is normalized by $6 N_s^2$, i.e., the surface of a cube with side length $N_s$.
In Fig.~\ref{fig:surface_vs_T} we show the surface of the largest cluster versus temperature for our $48^3 \times N_t$ lattices at all three values of the lattice constant. 
We observe that the surface is small below $T_c$, where all clusters are small. Near $T_c$, the largest cluster starts to percolate and 
also its surface increases drastically. For even higher temperatures, the cluster percolates and at the same time becomes denser by squeezing out the non-percolating clusters (compare Fig.~\ref{weight_np}),
and, thus, its surface slowly decreases as $T$ grows. Note that in this region ($T>T_c$) this observable suffers from large finite size effects, which are manifested in the 
difference between the three lattice spacings (corresponding to three different physical volumes).

\begin{figure}[t!]
 \centering
 \hspace*{-7mm}
 \includegraphics[width=75mm]{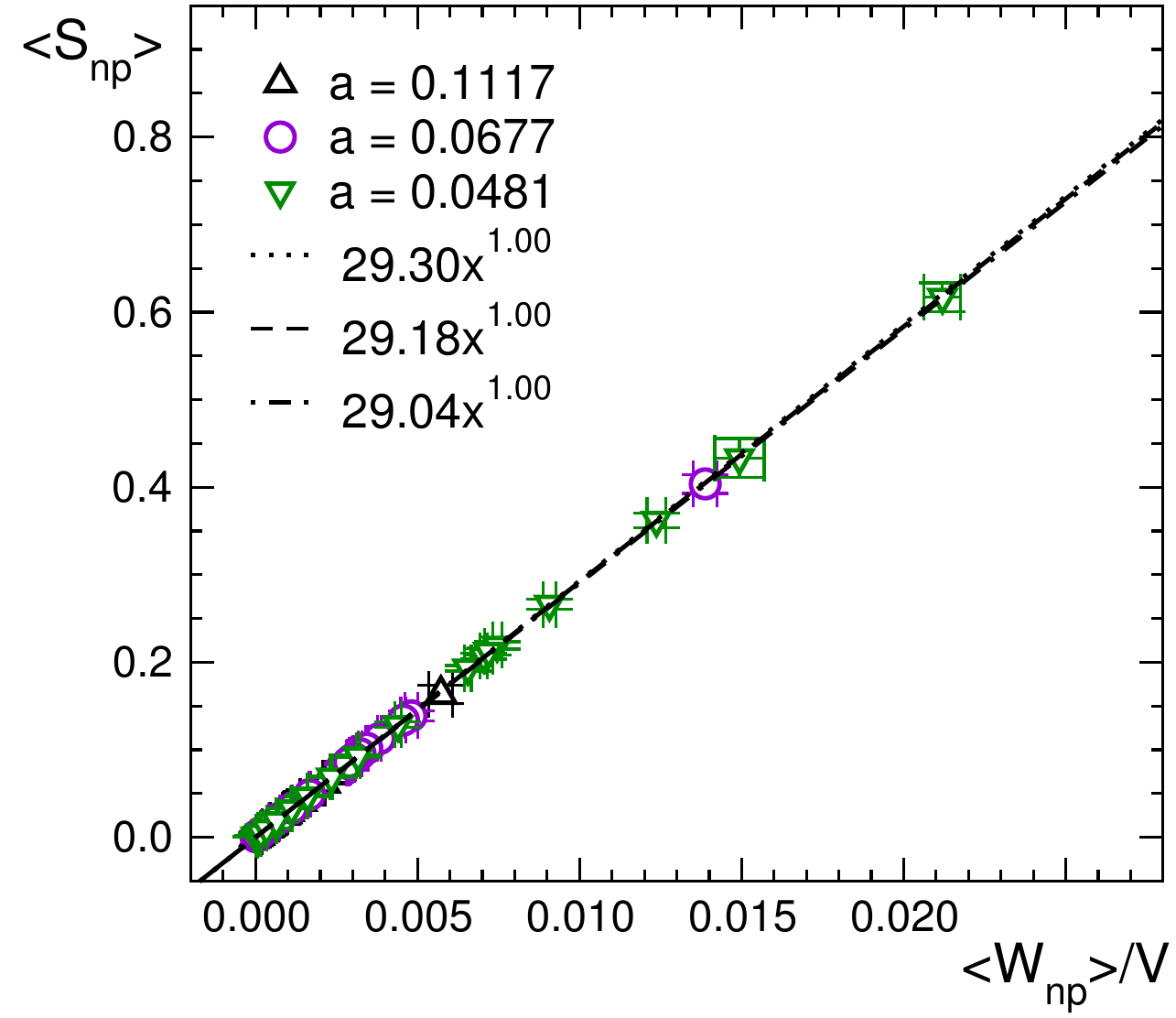}
 \hspace*{-7mm}
 \includegraphics[width=75.5mm]{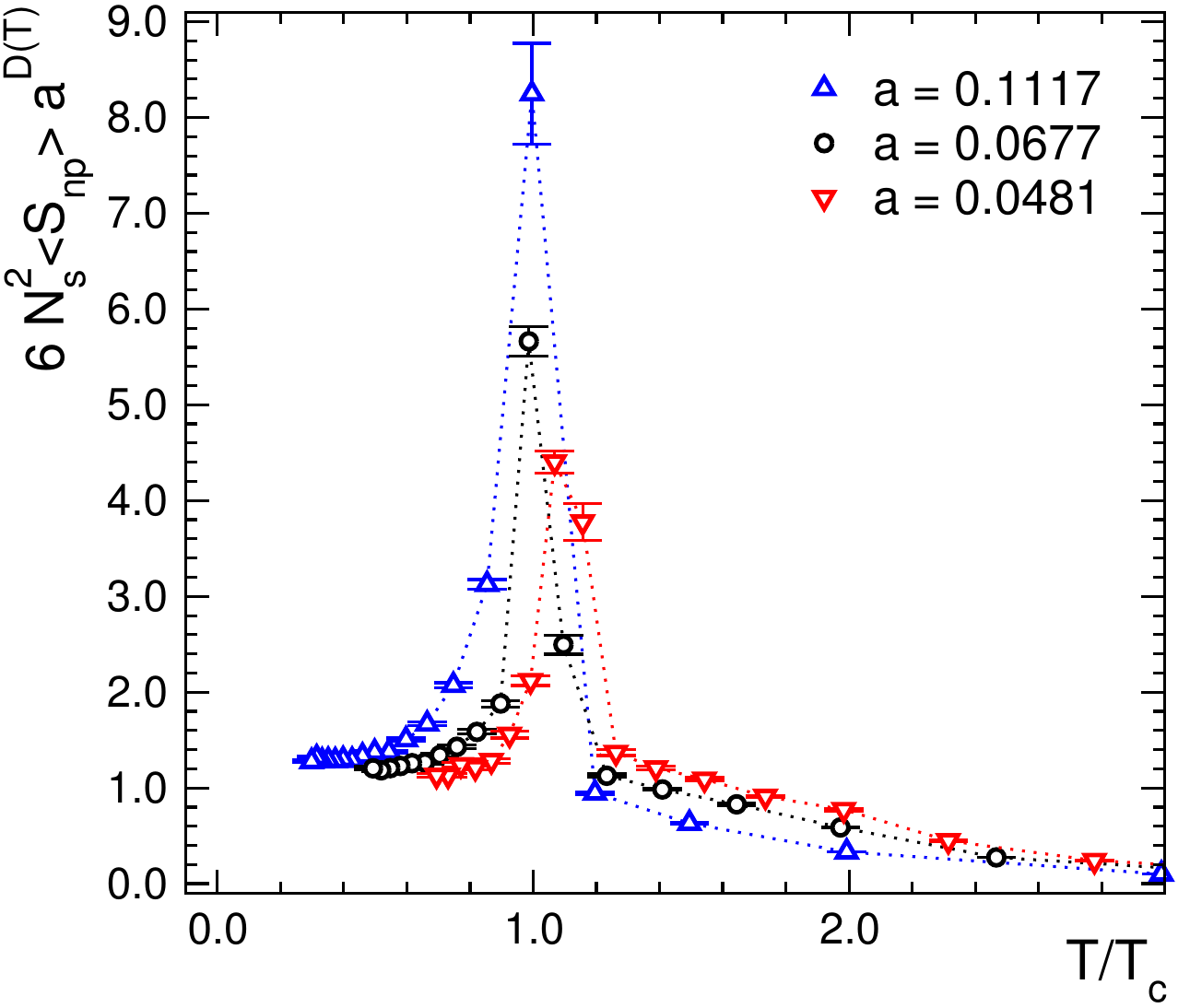}
 \caption{\label{fig:surface_vs_W} Upper panel: Average surface of the non-percolating clusters versus their average weight. The data are for our $48^3 \times N_t$ lattices using all $N_t$ and all three values of the lattice constant $a$. We also show the results of straight line fits of the data. Lower panel: The surface in physical units (for details see the text) as function of the temperature.}
\end{figure}

We proceed by considering the average surface $\langle S_{\rm np} \rangle$ of the largest non-percolating cluster. In the upper panel of Fig.~\ref{fig:surface_vs_W}, the surface is shown 
as function of the average weight $\langle W_{\rm np} \rangle$. 
The 
plot is for our $48^3 \times N_t$ lattices and combines all temperatures and the three lattice 
spacings. It is obvious that the data align on a straight line (the dotted and dashed lines in 
the plot are the results of linear fits). 
We have repeated the same analysis also for the largest clusters and found again a straight line behavior for temperatures below $T_c$. The percolating clusters above $T_c$ 
deviate from the linear behavior due to the finite size of the lattice. 
The observed linear behavior is characteristic for random 
percolation~\cite{stauffer1994introduction} and we again interpret 
this as an indication that the correlation introduced by the interaction 
is rather small.

The linear relationship between $S$ and $W$ also implies that the surface of the clusters 
inherits the fractal dimensionality, and scales with the same power of the cluster size as the weight. 
Our results for the fractal dimension of the largest non-percolating cluster therefore imply $S_{\rm np}\propto R^{D_{\rm sc,np}(T)}$ as the temperature is varied.
Using this relationship, we have access to the non-percolating cluster surfaces 
in physical units through a multiplication by $a^{D_{\rm sc,np}(T)}$. In the lower panel of Fig.~\ref{fig:surface_vs_W}, the rescaled surface is shown in the transition region for our three lattice spacings. 
This quantity behaves like a susceptibility and shows 
a very pronounced peak at $T_c$, where the surface increases quickly and then drops again above $T_c$, where the non-percolating clusters act as holes in the largest (percolating) cluster.

The analysis of the clusters therefore shows that the deconfinement transition is 
accompanied by a rapid increase of the surface of the center domains, but we remind the reader again, that these surfaces have a fractal nature and are certainly not smooth domain walls (such
a smooth behavior would be incompatible with a susceptibility-type behavior for the non-percolating clusters).
 
The fractal nature of the clusters also has implications for the mean free path $\lambda$ 
of a `test parton' that would traverse the lattice. In a small numerical test we considered a simplistic definition of $\lambda$ 
as the average distance between two cluster surfaces in directions parallel 
to the coordinate axes. 
We found that at low temperatures, the test parton experiences a mean free path 
even lower than the fixed radius $R_{\rm phys}$ of the clusters, due to their 
fractality. As the temperature increases and the non-percolating 
fractal clusters disappear, $\lambda$ eventually becomes -- together with the cluster 
radius -- equal to half the size of the lattice. 
In this mechanism the non-percolating clusters are expected to play the major role, 
since these are the ones that remain fractal objects in the whole temperature region.

\section{Summary and discussion}
 
In this paper we performed a comprehensive analysis of the center structure of strongly 
interacting matter in the vicinity of the deconfinement transition by studying the spatial 
distribution of the Polyakov loop.
We found that local clusters are formed where the phase of the Polyakov loop is in one of 
the three center sectors. Below $T_c$, these clusters are balanced in the sense that all 
three sectors are represented equally, implying that the total Polyakov loop averages to 
zero. On the other hand, above the critical temperature, one cluster starts to percolate, giving 
rise to a preferred center sector and a nonzero expectation value of the total Polyakov 
loop. This finding reproduces earlier results in the 
literature~\cite{quenched1,quenched2,full_prelim}. 

We remark at this point, that also for pure SU(2) gauge theory, where the deconfinement transition 
is of second order, the center clusters have been studied in detail in 
\cite{quenched1b,quenched2b}. Apart from the fact that 
there one only has two elements in the center group, clusters can be defined in a similar way
\cite{quenched1b} or using the same construction as here \cite{quenched2b}.
The percolation of the clusters at $T_c$ has been analyzed thoroughly and we refer the reader to
 \cite{quenched1b,quenched2b} for details. 

Here, we concentrated especially on the relationship between the radius, 
weight (volume) and surface of the 
center clusters, which revealed that the clusters are not three-dimensional objects 
but have fractal properties.
In particular, we found the largest cluster to increase its dimensionality from $\sim2.5$ 
to $3$ as the deconfinement transition is crossed and percolation occurs. We also discussed 
the largest non-percolating cluster, which turned out to keep its fractal dimension across 
the transition, dropping from $\sim 2.5$ to $\sim 2.1$ as $T$ is varied from $T=0.75\,T_c$ to $2.0\,T_c$.
The fractal nature of the clusters is also supported by the finding that their surface is 
proportional to their weight (instead of being proportional to the weight to the power $2/3$). 
These fractal properties were found to depend only mildly on the lattice spacing, which 
we take as a strong indication that the cluster structure of the QGP -- 
with isolated domains below $T_c$ and percolation above -- is a well-defined 
concept in the continuum limit.
We note moreover that most of the fractal characteristics are quite similar to those in random 
percolation theory (with the role of the occupation probability played by the temperature in 
our case), revealing that the correlation between neighboring Polyakov loops is weak. 

We would like to conclude by elaborating the idea put forward in Ref.~\cite{centermodels}, 
regarding the role of center domains in explaining certain characteristics of the behavior 
of the Quark Gluon Plasma in heavy-ion collisions. 

The first such characteristic is the small shear viscosity $\eta$ of the QGP. The shear 
viscosity is proportional to the mean free path $\lambda$ of partons according to kinetic 
theory. The cluster structure of the QGP can have a significant impact on $\lambda$, 
since the walls between clusters act as potential barriers and can thus reflect soft 
partons~\cite{centermodels}. 
The mean free path, however, does not equal the average cluster size, as was assumed in 
Ref.~\cite{centermodels}, since, as we have seen, 
the clusters are fractal objects. On the contrary, $\lambda$ can be given in terms of 
the density of walls, which can be translated to the surface of (non-percolating) 
clusters we discussed in 
Sec.~\ref{sec:surfaces}. This surface becomes very dense around $T_c$, 
where the largest cluster just started to 
percolate but is still pierced by smaller, fractal-like clusters as holes. 
This picture thus implies the QGP to have a low shear viscosity 
in the vicinity of the deconfinement phase transition, in line with the experimental 
observations regarding the ratio $\eta/s$.

\begin{figure}[t]
 \centering
 \vspace*{.2cm}
 \includegraphics[width=8cm]{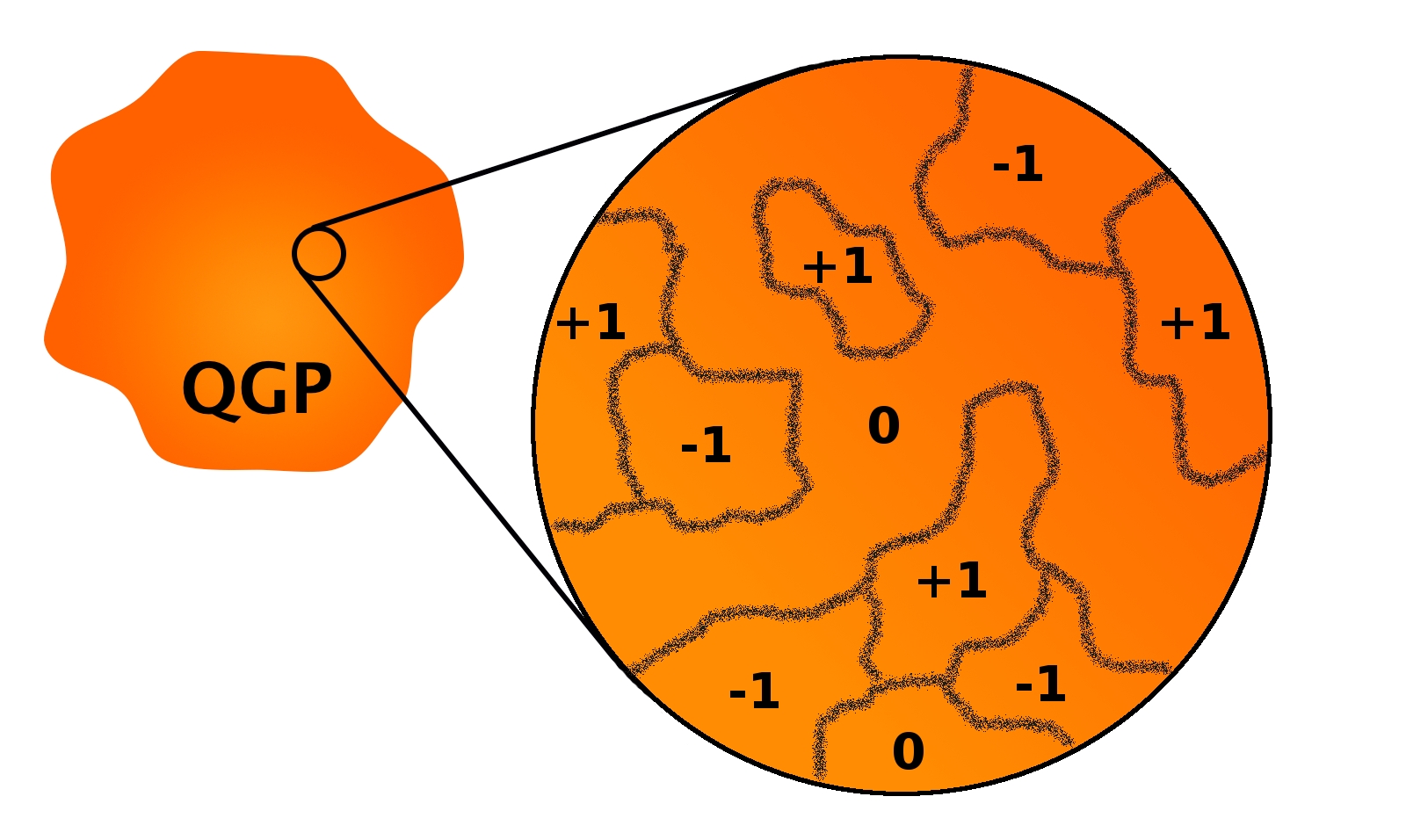}
 \caption{\label{fig:conclusion}Schematic illustration of center clusters slightly above $T_c$; figure adapted from Ref.~\cite{centermodels}.}
\end{figure}

A second consequence of the cluster structure is the large opacity of the QGP 
manifested via the quenching of high-energy jets: 
hard partons crossing cluster walls will emit soft gluons that 
further scatter on the walls, leading to a large energy loss~\cite{centermodels}. 
Using again the fact that the density of walls is high around $T_c$, 
this mechanism suggests that the QGP is very opaque for partons with high momenta in this temperature region. 
For both phenomena, we expect that the inclusion of dynamical fermions will somewhat 
increase the range in temperature, where the cluster structure is effective, 
due to the weakening (broadening) of the transition.

Note that our results indicate that the density of domain 
walls gradually decreases as the temperature grows and the 
fractal holes corresponding to the subdominant sectors diminish 
in size: Above $T_c$ the percolating cluster increases further in weight (Fig. 7) 
by becoming denser. Thus there are less sites available for the non-percolating 
clusters and thus their overall weight decreases when increasing $T$ 
further above $T_c$ (Fig. 8). According to the picture described above, the
decrease of the domain wall density implies 
a gradual increase in the QGP shear viscosity -- in qualitative 
agreement with recent experimental results comparing $\eta/s$ 
measured at RHIC and at the LHC. These measurements reveal 
an increase of the viscosity as $T$ grows~\cite{etaovers}, even 
though the $T$-dependence has not yet been fully established and 
is still under discussion, see, e.g., Ref~\cite{etaovers2}.
(Note that for even higher temperatures one expects the QGP to 
be well described as a gas of quasi-free partons, where the
mean free path and thus the shear viscosity are both large.)

A preliminary study of full QCD \cite{full_prelim} (which we currently repeat using the cleaner fixed scale approach) 
indicates that the explicit breaking from the fermion determinant is very weak and that the distribution 
properties of the local Polyakov loops are very similar to the quenched case, despite the fact that in full QCD one 
only has a crossover. Thus we expect that the physics 
implications of the domain walls are rather similar in the two cases. A conclusive answer will be possible only 
after the fixed scale approach full QCD studies are completed (which will be part two of the current paper). 

We illustrate our findings in Fig.~\ref{fig:conclusion}, which aims to depict the fractal 
structure of the QGP at a temperature slightly above $T_c$, where one percolating 
cluster (here in the center sector $0$) is filled by smaller, non-percolating 
cluster holes.
 
\vskip3mm
\noindent
{\bf Acknowledgements: } The authors would like to thank S.\ Bors\'anyi, J.\ Danzer,
M.\ Dirnberger, C.B.\ Lang, 
A.\ Maas, B. M\"uller, A. Sch\"afer and A.\ Schmidt for interesting discussions.  
This work is partly supported by DFG TR55, ``{\sl Hadron Properties from Lattice QCD}'' and 
by the Austrian Science Fund FWF Grant.\ Nr.\ I 1452-N27.
H.-P.~Schadler is funded by the FWF DK W1203 ``{\sl Hadrons in Vacuum, Nuclei and Stars}'' 
and a research grant from the Province of Styria. 
G.~Endr\H{o}di acknowledges support from the EU (ITN STRONGnet 238353) 
and from the Alexander von Humboldt Foundation.

\end{document}